\documentclass[12pt]{iopart}
\usepackage{geometry,graphicx, iopams,float,subfigure}

\expandafter\let\csname equation*\endcsname\relax

\expandafter\let\csname endequation*\endcsname\relax
\usepackage{amsmath}

\begin{document}

\title{The role of chaotic and ordered trajectories in establishing Born's rule}
\author{A.C. Tzemos\footnote{Corresponding Author} and G. Contopoulos}

\address{Research Center for Astronomy and 
Applied Mathematics of the Academy of 
Athens - Soranou Efessiou 4, GR-11527 Athens, Greece}
\ead{atzemos@academyofathens.gr,
gcontop@academyofathens.gr}
\vspace{10pt}

\begin{abstract}
We study in detail the trajectories, ordered and chaotic, of two entangled Bohmian qubits when their initial preparation satisfies (or not) Born's rule for various amounts of quantum entanglement. For any non zero value of entanglement ordered and chaotic trajectories coexist and the proportion of ordered trajectories increases with the decrease of the entanglement. In the extreme cases of zero and maximum entanglement we have only ordered and chaotic trajectories correspondingly. The chaotic trajectories of this model are ergodic, for any given value of entanglement, namely the limiting distribution of their points does not depend on their initial conditions. Consequently it is the ratio between ordered and chaotic trajectories which is responsible for the dynamical establishment (or not) of Born's rule.

\end{abstract}

\section{INTRODUCTION}
Bohmian Quantum Mechanics (BQM) is an alternative interpretation of Quantum Mechanics (QM) where the quantum particles follow certain deterministic trajectories guided by the usual wavefunction $\Psi$ (the solution of Schr\"{o}dinger's equation) according to the so called Bohmian equations (BE):
\begin{align}
m_i\frac{dr_i}{dt}=\hbar\Im\left(\frac{\nabla\Psi}{\Psi}\right)
\end{align}
BQM is a highly nonlocal theory where quantum entanglement has a strong impact on the evolution of quantum trajectories \cite{Bohm, BohmII, durr2009bohmian}. Moreover, since the BE are nonlinear one expects to find both ordered and chaotic trajectories.  There are many works  which study in detail and from many perspectives  both ordered and chaotic Bohmian  trajectories (\cite{iacomelli1996regular, frisk1997properties, falsaperla2003motion, wisniacki2005motion,efthymiopoulos2006chaos, wisniacki2007vortex, borondo2009dynamical}).  

In a series of previous works we focused on the production of chaos in Bohmian trajectories and came up with a generic theoretical mechanism for the emergence of chaos in arbitrary 2d and 3d systems \cite{efth2009, tzemos2018origin}. This is the so called nodal point-X-point complex (NPXPC) mechanism, which states that whenever a Bohmian particle comes close to the neighborhood of a moving nodal point of the wavefunction (the point where $\Psi=0$) it gets scattered by its accompanying X-point (a hyperbolic point of the Bohmian flow in the frame of reference of the moving nodal point with the same velocity with that of the nodal point).  The cumulative action of such close encounters between the particle and the NPXPCs leads to the emergence of chaos (for a review of chaos in BQM see \cite{efth2017, contopoulos2020chaos}).

In previous papers \cite{tzemos2019bohmian, tzemos2020chaos, tzemos2020ergodicity} we considered the problems of chaos, ergodicity  and applicability of Born's rule in a two qubit system, composed of coherent states of the quantum harmonic oscillator \cite{asboth2004coherent, garrison2008quantum}. This system has many interesting feautures that facilitate the study of many different aspects in BQM:
\begin{itemize}
\item Its entanglement can be calculated analytically since it is in close analogy with a two spin based qubit model (for entanglement in Bohmian trajectories see also \cite{zander2018revisiting, elsayed2018entangled}).
\item It has infinitely many NPXPCs lying on a straight lattice which moves and rotates in the configuration space. These lattices exist for every nonzero value of entanglement. The NPXPCs go to infinity at certain times and then they reappear. We found that all partially entangled states produce both chaotic and ordered trajectories, while in the two extreme cases of zero and maximum entaglement we have only ordered and only chaotic trajectories correspondingly. 
\item Its probability density $P=|\Psi|^2$ is characterized by two well defined blobs which move in the configuration space and collide close to the origin. During these collisions we see the temporary formation of several blobs between the nodal points of $\Psi$.  After a collision the two blobs are reformed and move on until the next collision and so forth.
\item Its ordered trajectories lie on a certain region of the configuration space, for given parameters.
\item The chaotic trajectories of our model were found to be ergodic for every given amount of the entanglement. Namely, their final distribution of points is the same regardless of their initial conditions (for ergodicity in Bohmian trajectories see also \cite{aharonov2004time, avanzini2017quantum}). 
\item The ergodic nature of the chaotic trajectories relies heavily on the existence of infinitely many NPXPCs. Consequently, although it is a simple quantum mechanical system, it exhibits very rich dynamics from a Bohmian standpoint. 
\end{itemize}

The above features make our model useful for the study of the origin of Born's rule, an important open problem in BQM \cite{valentini1991signalI, valentini1991signalII, durr1992quantum,  valentini2005dynamical, towler2011time, abraham2014long, durr2019typicality}.  Born's rule (BR) states that the probability density of finding a quantum particle in a certain region of space is equal to the absolute square of its wavefunction, namely:
\begin{align}
P=|\Psi|^2
\end{align}
It is well known in BQM that if Born's rule is initially satisfied, namely if the initial distribution of quantum particles $P_0$ is equal to $|\Psi_0|^2$, then it is satisfied for all times. However in BQM we can start with an arbitrary initial distribution with $P_0\neq |\Psi_0|^2$. Since Born's rule has
never been doubted by the experiment, we study the mechanism responsible for its dynamical establishment.

Most of our previous work was numerical. We found cases where Born's rule was established and cases where it was not established and we concluded that the amount of entanglement and the  nature of the trajectories in the distribution of Born's rule is responsible for its dynamical establishment, something that is true but not sufficient. 

In the present work we  study in detail with analytical formulae the mathematical background of our previous numerical results. Moreover we  provide further simulations in order to separate clearly the cases of the accessibility (or not) of Born's rule by an arbitrary initial distribution. The core result of our analysis is that the ergodicity of chaotic trajectories implies that an arbitrary initial distribution will finally come close to Born's rule distribution if the ratio between its  chaotic and ordered trajectories is close to that of the distribution satisfying BR.

In Section 2 we give the model of the two entangled qubits and in Section 3 we consider the time evolution of its probability density $|\Psi|^2$. In Section 4 we consider the nodal points of our model (where $\Psi=0$) and the evolution of the corresponding NPXPCs. We then study distributions of particles for various amounts of entanglement, when Born's rule is initially satisfied (Section 5) and the role of chaotic vs ordered trajectories in deriving (or not) Born's rule in the long run (Section 6). In Section 7 we find for what initial distributions of particles the final pattern is close to Born's rule.  Finally, in  section 8 we draw our conclusions. In the Appendix we present an approximative algorithm for the distinction between  the ordered and the  chaotic trajectories of our model.

\section{THE MODEL}

The case of 2 qubits deals with a most general solution of Schr\"{o}dinger's equation correspoding to a classical case of two harmonic oscillators, namely to a Hamiltonian of the form $
H=\frac{1}{2}(p_x^2/m_x+p_y^2/m_y+m_x\omega_x^2x^2+m_y\omega_yy^2)$. 
The solutions of the Schr\"{o}dinger equation are of the form
\begin{align}\label{Psigen}
\Psi=c_1Y_R(x,t)Y_L(y,t)+c_2Y_L(x,t)Y_R(y,t),
\end{align}
where
\begin{align}
\nonumber Y(x,t)=\Bigg(\frac{m_x\omega_x}{\pi\hbar}\Bigg)^{\frac{1}{4}}
\exp\Bigg[&-\frac{m_x\omega_x}{2\hbar}\Bigg(x-\sqrt{\frac{2\hbar}{m_x\omega_x}}
a_0\cos(\sigma_x-\omega_x t)\Bigg)^2\\&+i\Bigg(\sqrt{\frac{2m_x\omega_x}{\hbar}}
a_0\sin(\sigma_x-\omega_x t)x+\frac{1}{2}\Big[a_0^2
\sin(2(\omega_x t-\sigma_x))-\omega_x t\Big]\Bigg)\Bigg],
\end{align}
and the corresponding expression for $Y(y,t)$.

The entanglement depends on the values of $c_1$ and $c_2$ ($|c_1|^2+|c_2|^2=1$). In particular if $c_2=0$ we have a product state with no entanglement. 
We work with $\hbar=m_x=m_y=1, \omega_x=1,\omega_y=\sqrt{3}$ and $a_0=5/2$. Moreover $\sigma_x=\sigma_y=0$ for $Y_R$, while $\sigma_x=\sigma_y=\pi$ for $Y_L$. The values of $\omega_x, \omega_y$ have a non commensurable ratio, while $a_0$ is sufficiently large in order to secure the qubit character of the solution. 
Thus
\begin{align}
Y_R(x,t)\!=\!\Big(\frac{\omega_x}{\pi}\Big)^{\frac{1}{4}}\exp\!\Bigg[\!-\frac{\omega_x}{2}\left(x-\sqrt{\frac{2}{\omega_x}}a_0\cos(\omega_xt)\right)^2+i\! \left( -\sqrt {2\omega_x}a_{0}x\sin \left( \omega_{x
}t \right) +\frac{a_{0}^{2}\sin( 2\omega_{x}t)  -\omega
_{x}t} {2}\right) \!
    \Bigg]
\end{align}
while in $Y_L(x,t)$ the factor in the square inside the exponent is $[x+\sqrt{\frac{2}{\omega_x}}a_0\cos(\omega_x t)]$. The terms $Y_R(y,t)$ and $Y_L(y,t)$ are similar.

\begin{figure}[H]
\centering
\includegraphics[scale=0.17]{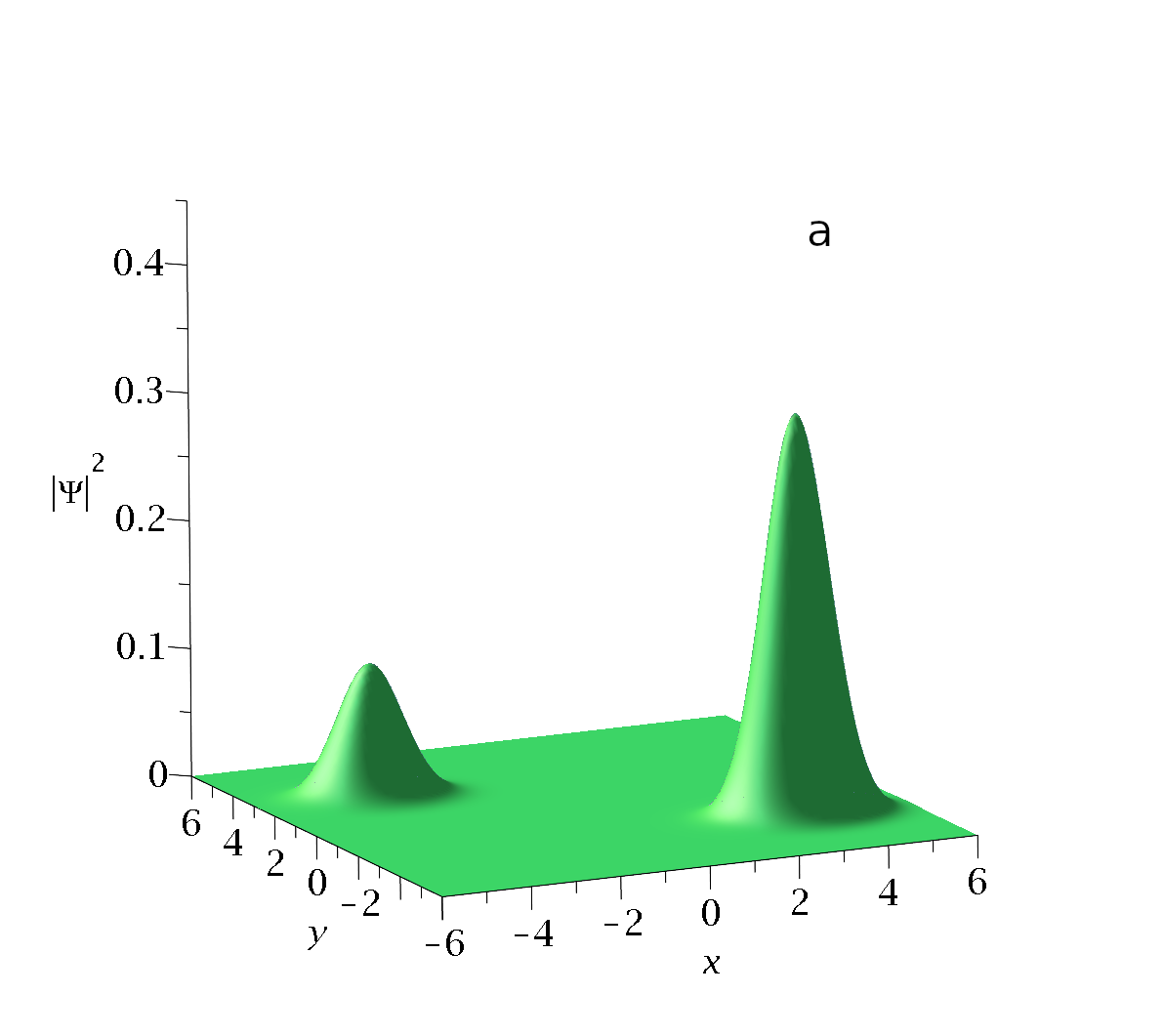}
\includegraphics[scale=0.17]{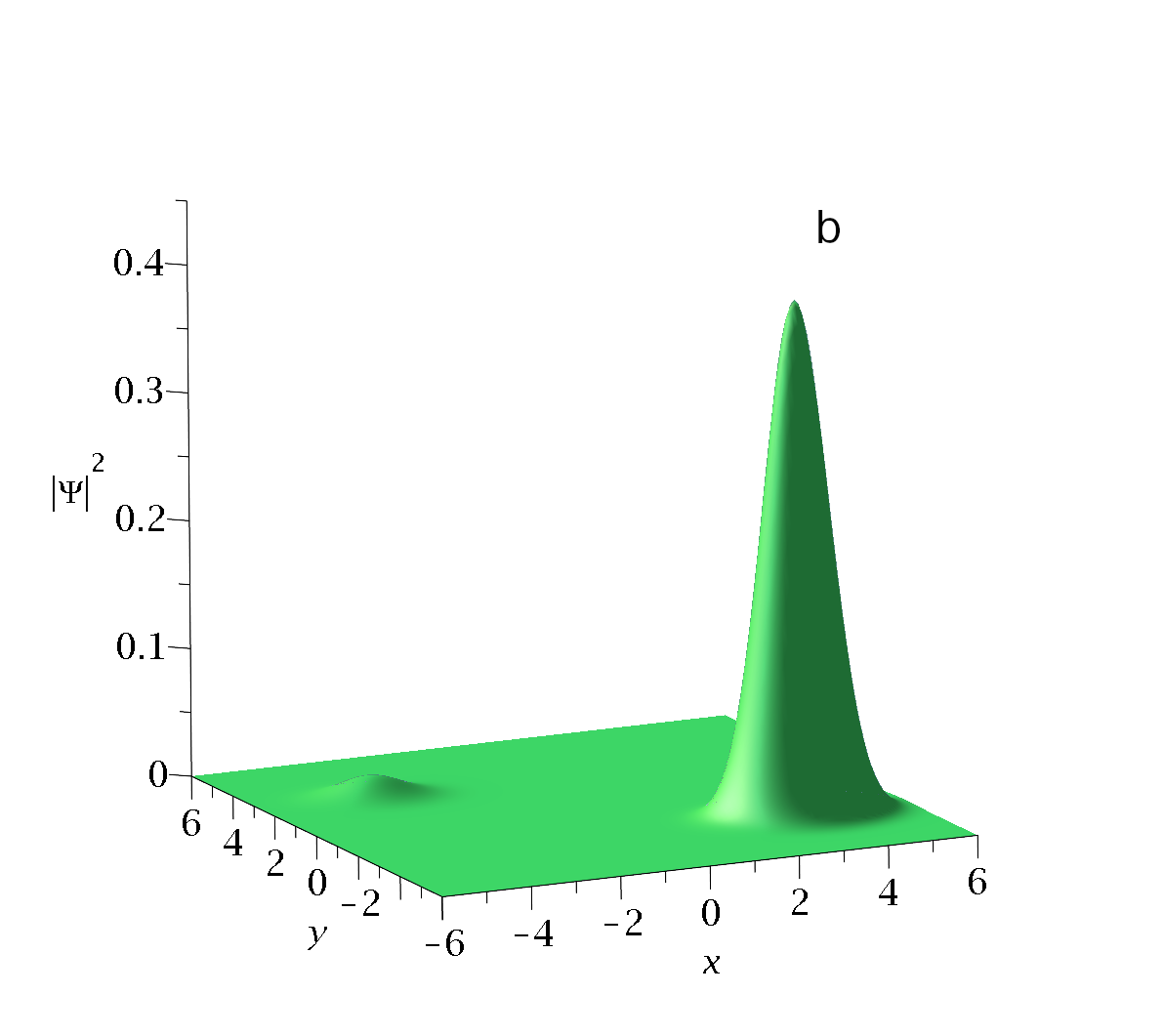}
\caption{The initial form of $|\Psi|^2$ in the case (a) $c_2=0.5$  and (b) $c_2=0.2$. }\label{t0}
\end{figure}

\begin{figure}[H]
\centering
\includegraphics[scale=0.37]{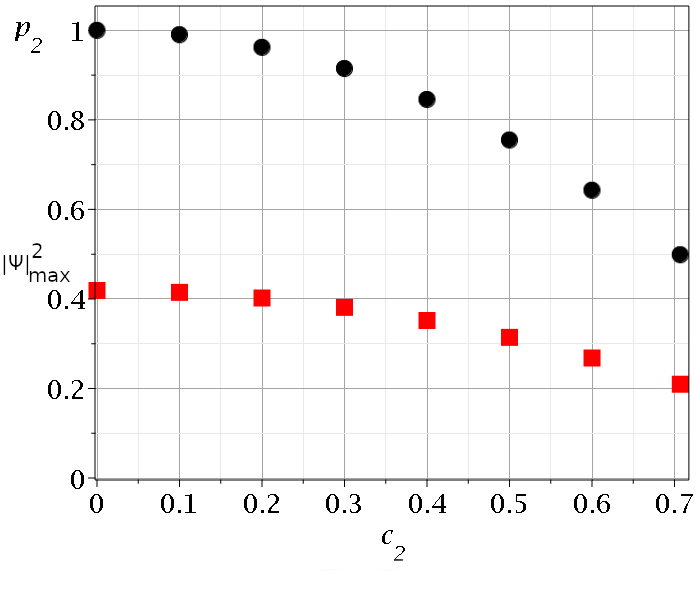}
\caption{The maximum height of the main blob of $|\Psi_0|^2$ (red squares) and the percentage of the particles on the main blob $p_2$ (black dots) as   functions of the entanglement parameter $c_2$.}\label{perc2}
\end{figure}

For any  non zero value of the entanglement the initial distribution $|\Psi_0|^2$ consists of two Gaussian blobs, one on the lower right quadrant and one on the upper left quadrant of the configuration space. We consider mainly cases where $c_2<c_1$ in which the first blob is larger (Figs.~\ref{t0}a,b for the cases $c_2=0.5$ and $c_2=0.2$). In the case $c_2=c_1=\sqrt{2}/2$ the two blobs are equal (Fig. 2 of our paper \cite{tzemos2020ergodicity}). If $c_2=0.5$ the maximum height of the secondary blob is about $1/3$ of the main blob and if $c_2=0.2$ it is only $0.04$ of the main blob. In the latter case the secondary blob is barely seen in Fig.~\ref{t0}b. The value of the maximum $|\Psi_0|^2$ as a function of $c_2$ is given in Fig.~\ref{perc2}. The volume of the main blob (which gives the proportion of the particles of this blob $p_2$) is also given as function of $c_2$ in Fig.~\ref{perc2}. The ratio $p_1/p_2$ is very close to the corresponding ratio between the maximum heights of the two blobs. If $c_2=0$ there is only one blob.

\section{TIME EVOLUTION OF THE PROBABILITY DENSITY $|\Psi|^2$}

 The values of $|\Psi|^2$ in the product state ($c_2=0$) form a blob around a center given by
\begin{align}\label{xc1yc1}
x_c=\sqrt{\frac{2}{\omega_x}}a_0\cos(\omega_x t), y_c=-\sqrt{\frac{2}{\omega_y}}a_0\cos(\omega_y t), 
\end{align}
For these values of $x,y$ we have
%
%
\begin{align}
\Psi=Y_R(x_c,t)Y_L(y_c,t)=\frac {(\omega_{x}\omega_y)^{\frac{1}{4}}}{\sqrt {\pi}}{ \exp\left[-\frac{i}{2} \left( {a_{0}}^{2}\left(\sin \left(2 \omega_{x}t \right)+\sin \left( 2\omega_{y
}t \right)\right)+{(\omega_{x}+\omega_y)}t \right) \right]}\label{Psig}
\end{align}
and 
\begin{align}\label{Psimax}
|\Psi|^2_{max}=\frac{\sqrt{\omega_x\omega_y}}{\pi}.
\end{align}

As a consequence the blob follows a Lissajous figure given by Eqs.~(\ref{xc1yc1}) with a constant amplitude, given by Eq.~(\ref{Psimax}). For $t=0$ the position of the center of the blob is at $x_c=\sqrt{\frac{2}{\omega_x}}a_0,  y_c=-\sqrt{\frac{2}{\omega_y}}a_0$  
and for the above values of $\omega_x, \omega_y$ and $a_0$ it is $(x_c'=3.54,y_c'=-2.69)$. A symmetric solution occurs in the product state with $c_1=0$. In that case there is a blob which is initially around the point $(x_c=-3.54, y=2.69)$ and forms a Lissajous figure 
\begin{align}
x_c'=-\sqrt{\frac{2}{\omega_x}}a_0\cos(\omega_x t),\quad
y_c'=\sqrt{\frac{2}{\omega_y}}a_0\cos(\omega_y t),\label{xcycd}
\end{align}
symmetric with respect to the trajectory (\ref{xc1yc1}).

However, when $c_2\neq 0$ the situation is more complicated. If we set the solution (\ref{xcycd}) in Eq.~(\ref{Psigen}) the term $c_1Y_R(x,t)Y_L(y,t)$ is $c_1$ times the function $\Psi$ of Eq.~(\ref{Psig}), where the exponent has only an imaginary quantity. But we have also the term $c_2Y_L(x,t)Y_R(y,t)$, which has a real exponential besides the imaginary part. This term is

\begin{align}
\nonumber c_2Y_L(x,t)Y_R(y,t)=c_2\frac{\sqrt{\omega_x\omega_y}}{\pi}\exp\left[-4a_0^2\left(\cos^2(\omega_x t)+\cos^2(\omega_y t)\right)\right]\\ \times
\exp\left[\frac{i}{2}\left({3}a_0^2\left(\sin(2\omega_x t)+\sin(2\omega_y t)\right)-\left(\omega_x+\omega_y \right)t\right)\right]\label{term}
\end{align}

The first two terms of the exponent are 
\begin{align}
E=\exp\Big[-25(\cos^2t+\cos^2(\sqrt{3}t)\Big]
\end{align}
when $\omega_x=1, \omega_y=\sqrt{3}, a_0=5/2$
then $E$ is larger than $\exp(-50)$.
In general this term is very small. Therefore the trajectory  of
the blob that is initially around $(x_c=3.54, y_c=-2.69)$, is very  close to the trajectory of the blob of the product state $c_2=0$. The other
blob is initially around the point ($x_c=-3.54, y_c=2.69$) and
forms an trajectory almost symmetric, with respect to the origin, to
that of the first blob. (The two trajectories are exactly symmetric if 
$c_1=c_2$).
Sometimes the quantity $(\cos^2t+\cos^2(\sqrt{3}t)$ is very
small and then the exponential factor becomes of order 1. This happens if $t$ is close to $k\pi/2$ and at the same time
close to $k_2\pi/(2\sqrt{3})$ for odd integers $k_1$ and $k_2$.
E.g. if $k_1=3  (t_1=4.71), k_2=5 (t_2=4.59)$ then $E$ is maximum $E=0.31$ between $t_1$ and $t_2$ for $t=4.58$. Similarly for $k_1=5, k_2=7$ we find a maximum $E=0.03$ for $t=8.1$. At the times $t=4.58$ and $t=8.1$ we have collisions of the two blobs because their distances from the origin is very small (see Fig.~\ref{nodal_nodal}). The distances of the top of one blob from the origin is:
\begin{align}
\Delta_o=\sqrt{\frac{2}{\omega_x\omega_y}}a_0\sqrt{\omega_y\cos^2(\omega_xt)+\omega_x\cos^2(\omega_yt)}
\end{align}

The collisions occur when the two blobs approach each other as they come close to the origin with their tops  forming almost symmetric Lissajous curves. In fact the tops of the blobs appear at the values of x and y where
\begin{align}\label{partial}
\frac{\partial |\Psi|^2}{\partial x}=\frac{\partial |\Psi|^2}{\partial y}=0
\end{align}

If we set the values (\ref{xc1yc1}) in $|\Psi|^2$ we find that in 
general the Eqs.~(\ref{partial}) are satisfied with very high accuracy, except for the times close to the collisions of the blobs. The collision times are approximately the same for any value of $c_2$, since the time interval  where the absolute values of the derivatives of $|\Psi|^2$ are larger than a small value of order $10^{-4}$ is about $\Delta t\simeq 0.5$.
Between collisions the two blobs form slightly  deformed Lissajous figures, therefore they stay longer at the four corners of these curves. 


\section{NODAL POINTS}

The wavefunction vanishes at the nodal points where  $\Psi_R=\Psi_I=0$. 
In this model we have an infinity of nodal points given by 
the formulae:
\begin{eqnarray}
\nonumber\label{xnod}&x_{nod}={\frac {\sqrt {2}
\left( k\pi\,\cos \left( 
\omega_{y}\,t \right) +\sin \left( 
\omega_{y}\,t \right) \ln  \left( 
\left| {\frac {c_{1}}{c_{2}}} \right|  
\right)  \right) }{4\sqrt {\omega_{x}}a_{0}\,\sin \left(  
\omega_{xy}  t \right) } }\\&
\label{ynod}y_{nod}={\frac {\sqrt {
2} \left(k\pi\, \cos \left( \omega_{x}t 
\right) +\sin \left( \omega_{x}t \right) 
\ln  \left(  \left| 
{\frac {c_{1}}{c_{2}}} \right|  \right)  
\right) }{4\sqrt {\omega_{y}}a_{0}\,\sin 
\left( \omega_{xy}\,t \right) }}
\end{eqnarray}
with $k\in Z $, $k$ even for $c_1c_2<0$
or odd for $c_1c_2>0$ and $\omega_{xy}\equiv \omega_x-\omega_y$. 
The differences between successive nodes $k-2$ and $k$ are:
\begin{align}
\Delta x={\frac {\sqrt {2}\pi\,\cos \left( \omega_{y}\,t \right) }{2\sqrt {
\omega_{x}}a_{0}\,\sin \left( \omega_{xy}t\right) }},\quad
\Delta y={\frac {\sqrt {2}\pi\,\cos \left( \omega_{x}\,t \right) }{2\sqrt {
\omega_{y}}a_{0}\,\sin \left( \omega_{xy}t\right) }}
\end{align}
Therefore the nodal points lie on a straight line with inclination
\begin{align}
\frac{\Delta y}{\Delta x}=\sqrt{\frac{\omega_x}{\omega_y}}\frac{\cos\omega_xt)}{\cos(\omega_yt)}
\end{align}

\begin{figure}[H]
\centering
\includegraphics[scale=0.25]{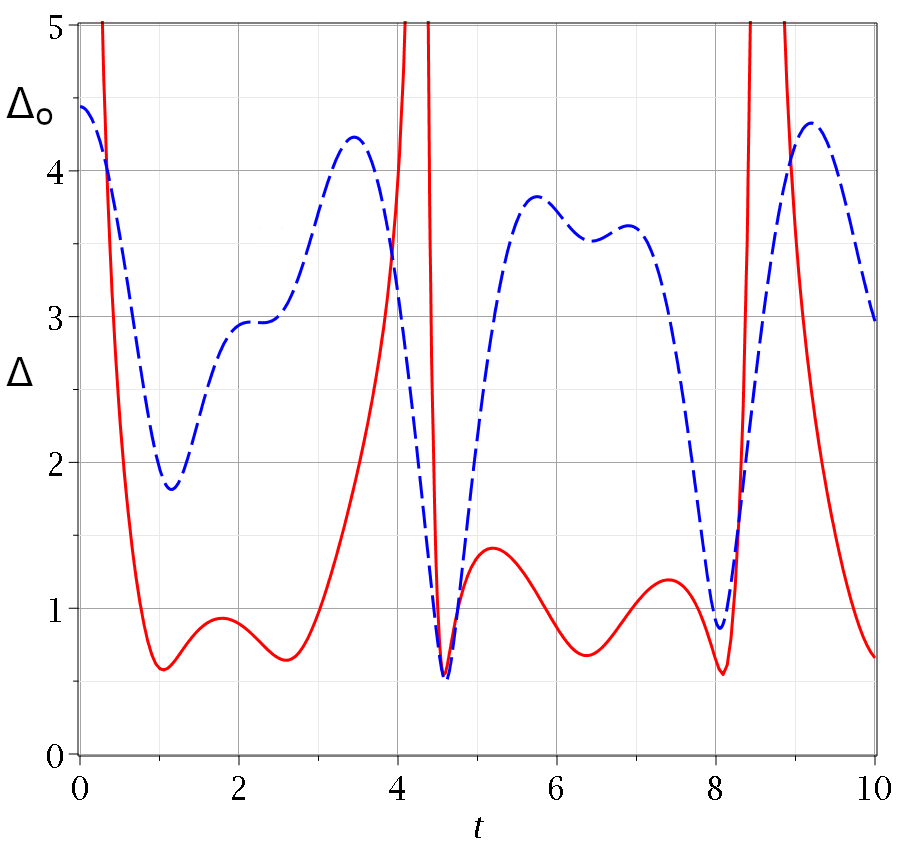}
\caption{The distance $\Delta_{o}$ between the center of the lower left blob and the origin for $t\in[0,10]$ (blue dashed curve) and the distance $\Delta$ between two successive nodal points ($k=-1 $ and $k=1$) (red curve). This distance is the same between any two consecutive nodal points and for any $c_2\neq0$.  }\label{nodal_nodal}
\end{figure}

At any time $t$ the distance between the node $k$ from the 
node $k-2$ is
\begin{align}
\Delta=\frac{\pi}{a_0|\sin(\omega_{xy}t)|}\sqrt{\frac{\omega_x\cos^2(\omega_xt)+\omega_y\cos^2(\omega_yt)}{2\omega_x\omega_y}}
\end{align}
It is of interest to note that this distance between successive nodes is the same for all $c_2\neq 0$ at the same time (Fig.~\ref{nodal_nodal}).
For $t=\frac{\Lambda\pi}{\omega_{xy}}$ with integer $\Lambda$, the nodes are at infinity and for an interval of $t$ their distances are given in Fig.~\ref{nodal_nodal}.

The line of nodes is at a distance:
\begin{align}
d_{no}=\frac{\sqrt {2}  {\ln  \left(  \left| {\frac {c_{1}}{c_{2}}}
 \right|  \right) } }{{4a_0}\sqrt {{ \left( \cos
 \left( \omega_{x}\,t \right)  \right) ^{2}\omega_{x}+\omega_{y}\,
 \left( \cos \left( \omega_{y}\,t \right)  \right) ^{2}}}}
\end{align}
from the origin. This distance depends on $t$ and $c_2$. In the particular case $c_1=c_2=\sqrt{2}/2$ this distance is zero, i.e. the line of nodes passes always through the origin and when $c_1\neq c_2$  this distance is larger than $d_{min}=\frac{\sqrt{2}\ln\Big(\Big|\frac{c_1}{c_2}\Big|\Big)}{4\sqrt{\omega_x+\omega_y}}\simeq0.086\ln\left(\left|\frac{c_1}{c_2}\right|\right)$. The line of nodes rotates clockwise and counterclockwise from time to time, thus covering most areas of the configuration space (see Fig.~1 of \cite{tzemos2020chaos}).
When the blobs of $|\Psi|^2$ are far from the line of nodes the value of $|\Psi|^2$ between the nodes is very small (less than $10^{-11}$).


\begin{figure}[H]
\centering
\includegraphics[scale=0.17]{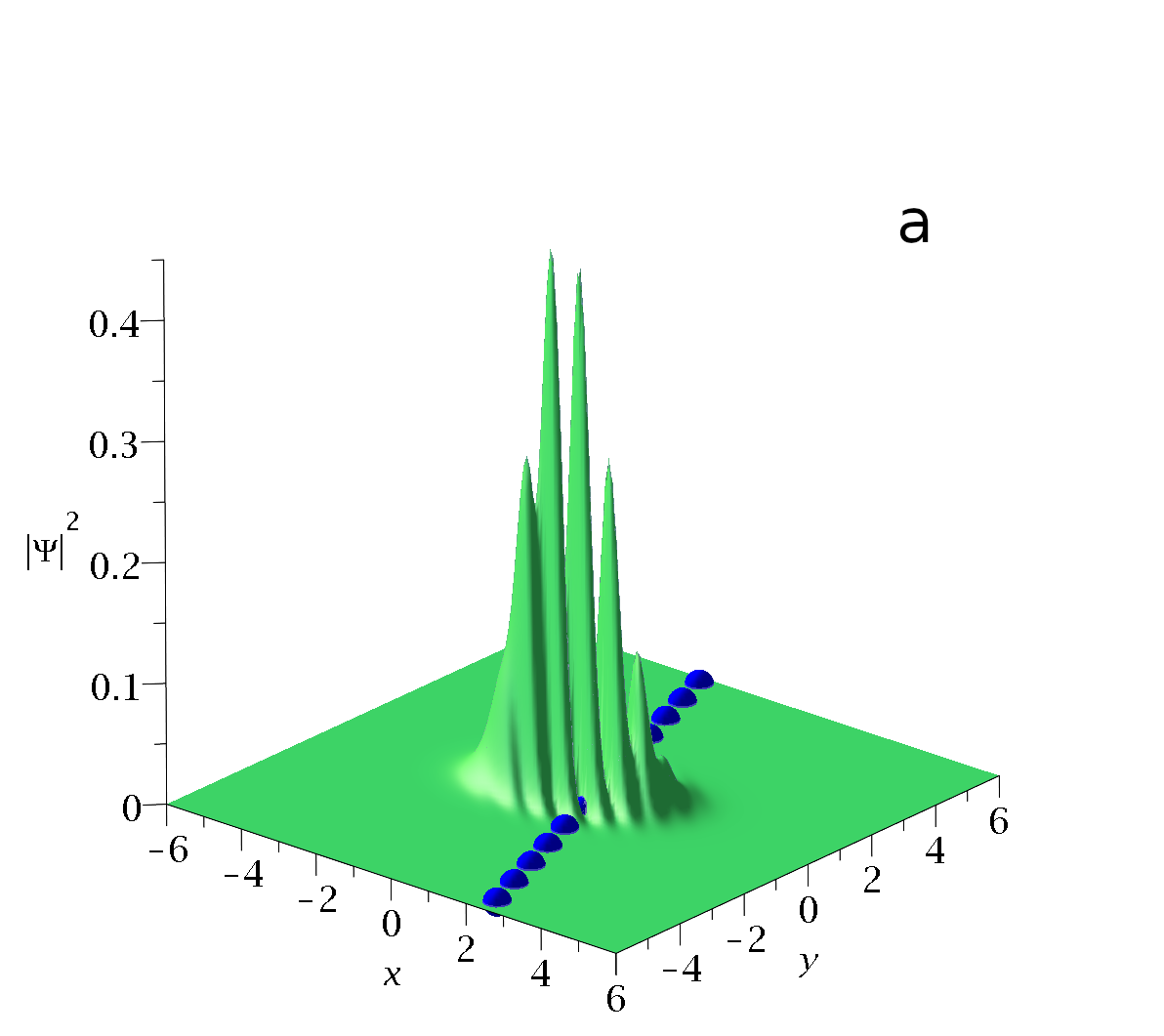}
\includegraphics[scale=0.17]{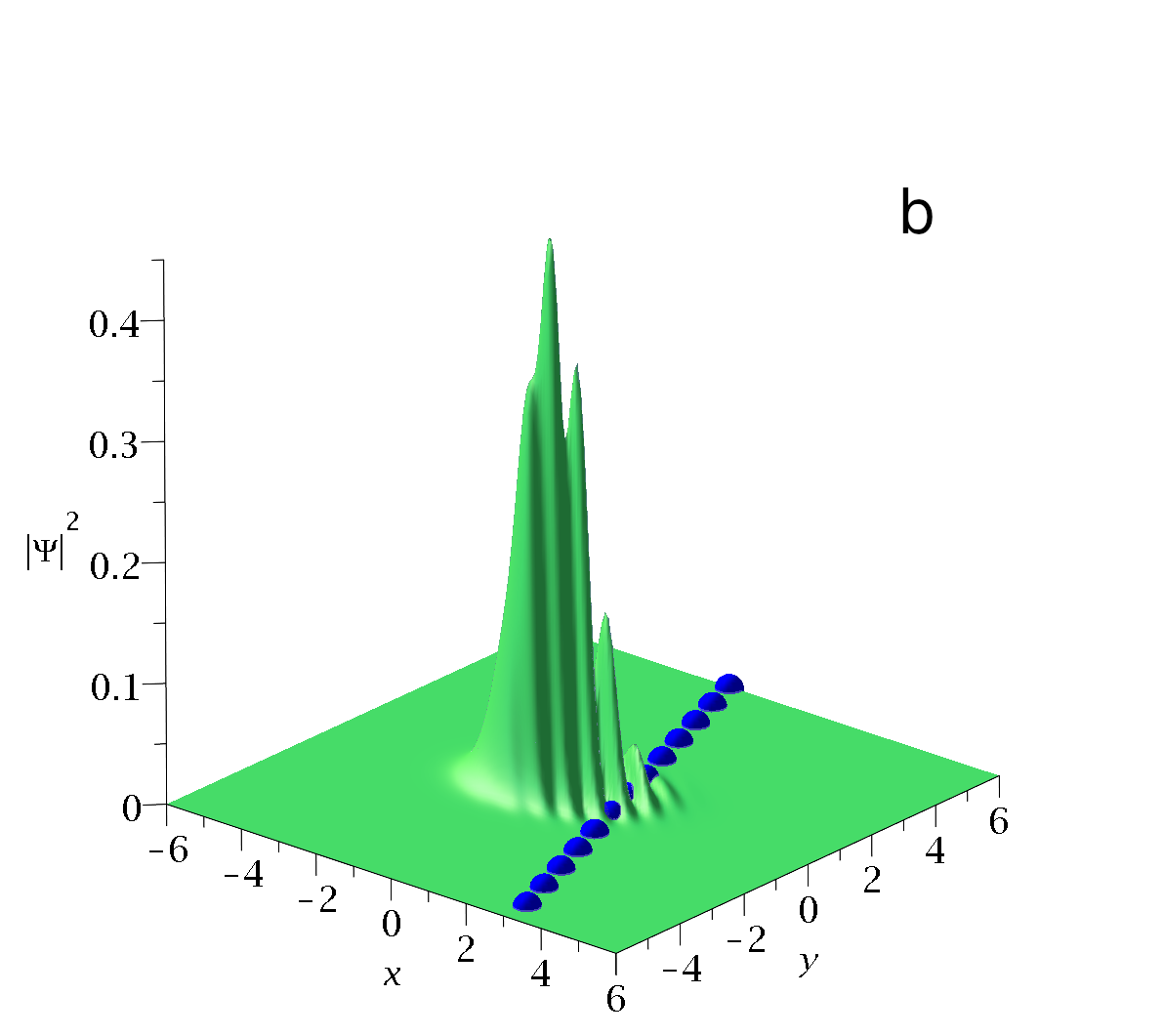}\\
\includegraphics[scale=0.17]{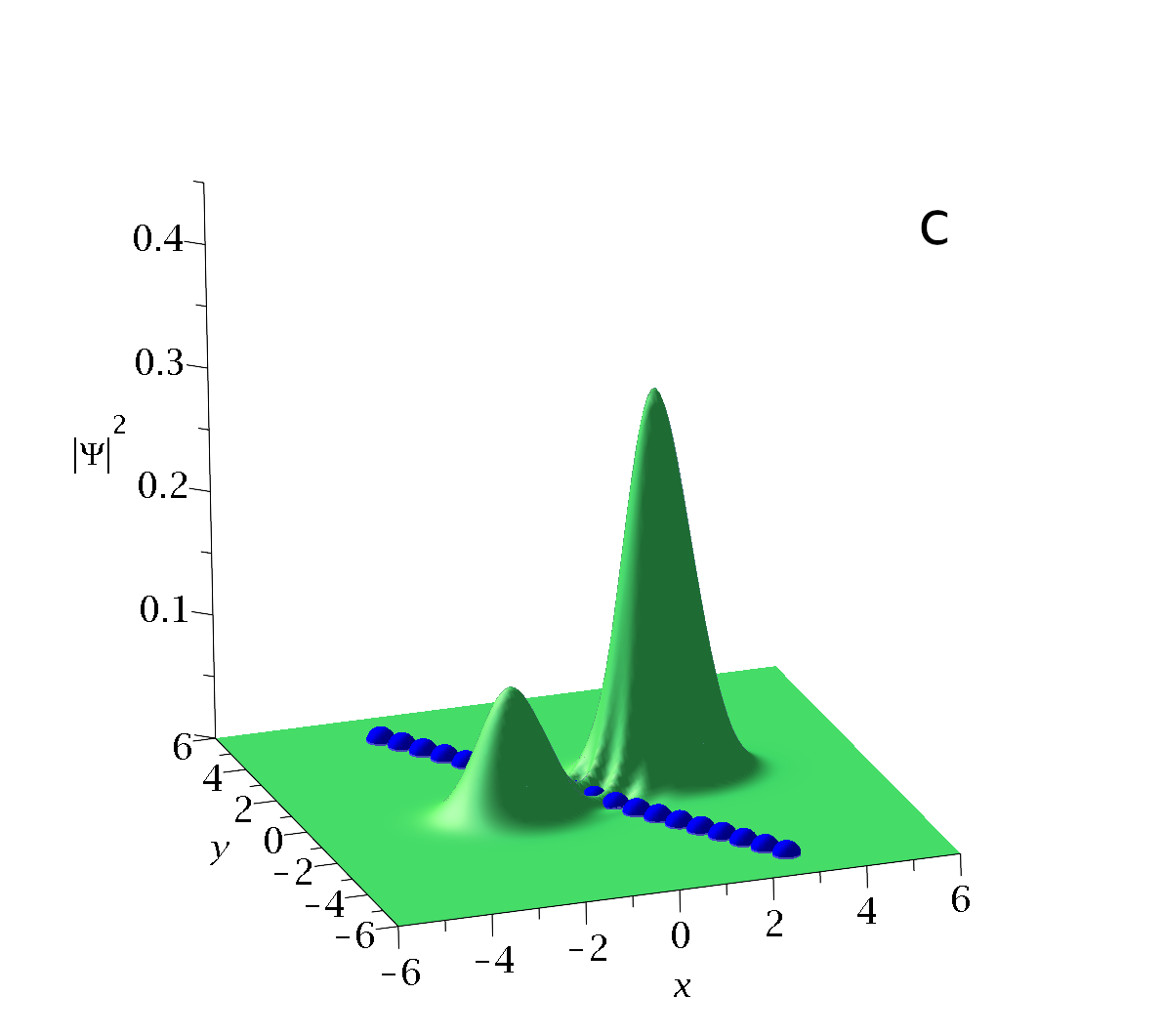}
\includegraphics[scale=0.17]{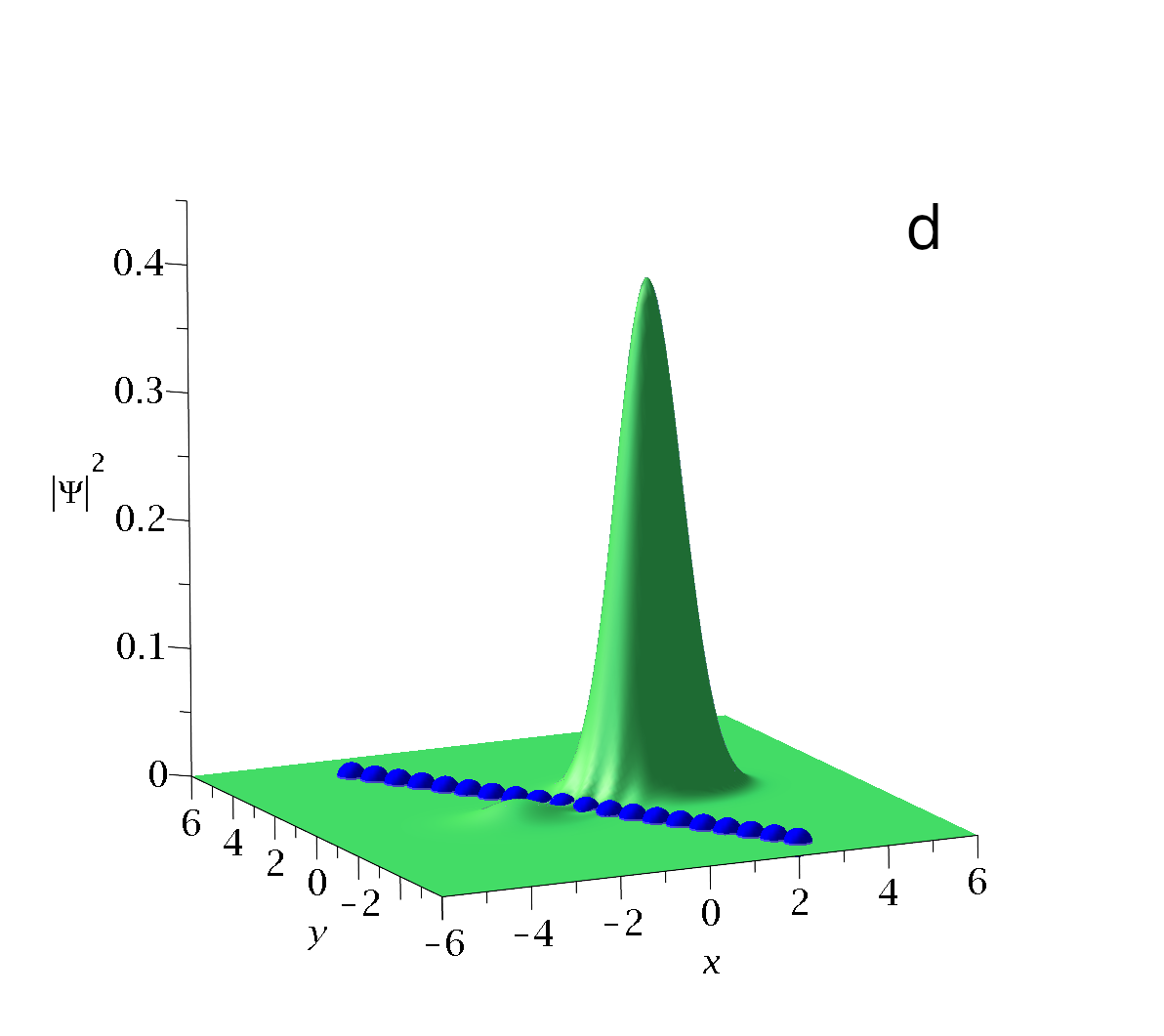}
\caption{3d plots of $|\Psi|^2$. Left column: $c_2=0.5$ for $t=4.5$ and $t=1.05$. Right column: $c_2=0.2$ for $t=4.5$ and $t=1.05$.}\label{nodal_origin}
\end{figure}

When the blobs approach the line of nodes the blobs are split into a number of secondary blobs between the nodes that are close to the origin  (Figs.~\ref{nodal_origin}ab).

This was seen in the case $c_2=\sqrt{2}/2$ (Fig.~2 of \cite{tzemos2020ergodicity}), where the two blobs are equal and  there are symmetric peaks on both sides of the origin at the peak of the collision. Here we show in Fig.~\ref{nodal_origin} the collisions of the blobs in the cases $c_2=0.5$ and $c_2=0.2$ where the two blobs are not equal in size.


At the collision (e.g. at $t=4.58$ ) the splittings are quite asymmetric (Figs.~\ref{nodal_origin}ab). The positions of the nodes are then at their closest distance from each other.

However at some minima of the distances between nodes we do not have collisions. E.g. this happens at the minima $t=1, t=3.2, t=6.3$ etc (Fig.~\ref{nodal_nodal}). In these cases the two blobs do not approach each other very close. Then only their outer parts may overlap (Figs.~\ref{nodal_origin}cd). 
In such cases only few particles of the distribution that correspond to the blobs are deflected (see section 5).

\begin{figure}[H]
\centering
\includegraphics[scale=0.27]{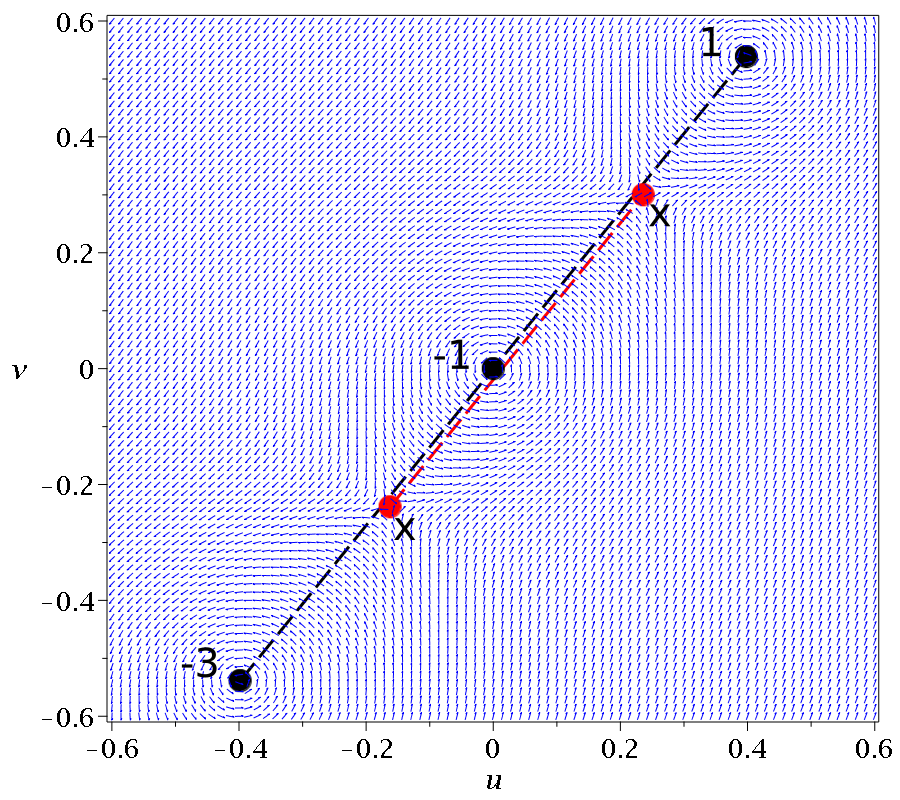}
\caption{The flow (small blue arrows) around the central nodal point $k=-1$ for $c_2=\sqrt{2}/2$ and $t=2.46$. The black dashed line joining the nodal points shows the direction of the nodal lattice at the current time. The X-points (red dots joined by the red dashed line) are very close to the black line.}\label{xp}
\end{figure}

The motion of the nodes dictates the motion of the NPXPCs, i.e. the characteristic structures of the Bohmian flow which are responsible for the generation of chaos \cite{efth2009}. Consequently, in order to monitor the scattering events underwent by the particles of a distribution,  one needs, besides the nodes, to mark also the position of the X-points. The X-points  are stationary in the frame centered at a moving nodal point and  deflect the approaching particles. An example of the lattice of the NPXPCs is shown in Fig.~\ref{xp}, where  we see that the X-points are about halfway between the nodes and very close to the line of nodes.

\section{DISTRIBUTIONS OF TRAJECTORIES WITH $P_0=|\Psi_0|^2$}

\begin{figure}[H]
\centering
\includegraphics[scale=0.3]{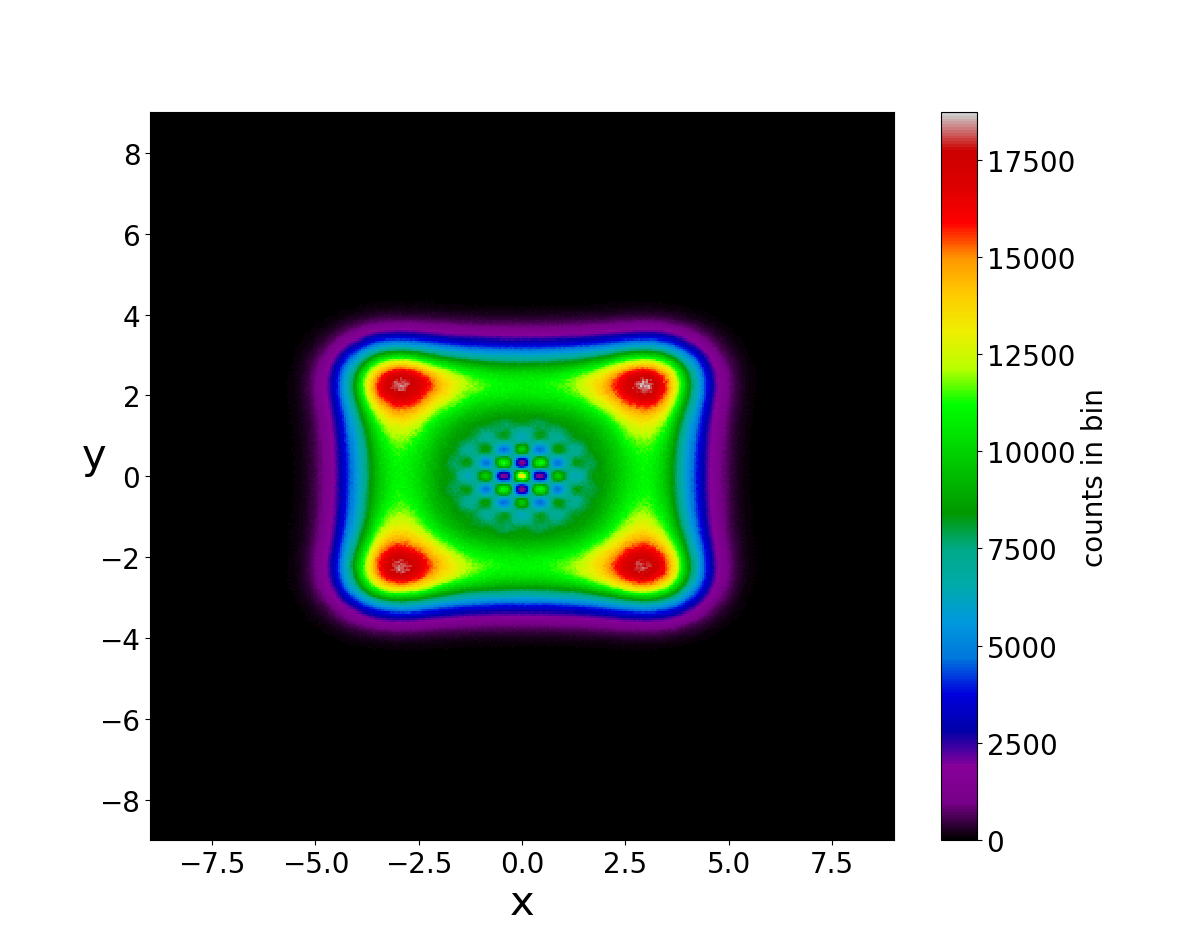}
\caption{Multiparticle distribution of 2400 particles in the case $c_2=\sqrt{2}/2$  when Born's rule is initially satisfied for times up to $t=5000$. }\label{0707_md_satisfaction}
\end{figure}

In our previous papers \cite{tzemos2020chaos,tzemos2020ergodicity} we considered the trajectories in the cases $c_2=\sqrt{2}/2\simeq 0.707$ (maximally entangled state), $c_2=0.5$ (strongly entangled state), $c_2=0.2$ (weakly entangled state) and $c_2=0$ (product state) and checked whether a distribution reaches the Born rule in the long run, by comparing the final pattern of the points of the trajectories of the initial distribution with that of the Born rule. These patterns are formed by collecting all the points of the trajectories inside the cells of a $360\times 360$ grid covering the space $[x,y]\in[-9,9]$ at times equal  to $t=n\Delta t (n=0, 1, 2, ...)$ and up  to a sufficiently large time $t_f$, with a step $\Delta t=0.05$ and plotting them by use of a spectral color plot.\footnote{We have checked that the patterns do not change if we take smaller values of $\Delta t$.} An example of such a pattern, with $t_f=5000$, is shown in Fig.~\ref{0707_md_satisfaction}. We also found  that  if  $P_0=|\Psi_0|^2$, then the long term distributions of the points of the trajectories form very similar patterns, like that of Fig.~\ref{0707_md_satisfaction}. 

As $t$ increases the patterns for any given $c_2$ tend to a final form. The evolution of the distributions  over the course of time and the differences between them can be studied using a matrix norm. In the present paper we work with the Frobenius norm $D$.\footnote{The Frobenius norm gives the distance $D$ between two matrices $A$ and $B$ according to the formula:
\begin{align}
D\equiv||A-B||=\sqrt{tr(A-B)^{\dagger}(A-B)}
\end{align} The details of an application of this norm in our particular problem are given in \cite{tzemos2020ergodicity}.}

\begin{figure}[H]
\centering
\includegraphics[scale=0.25]{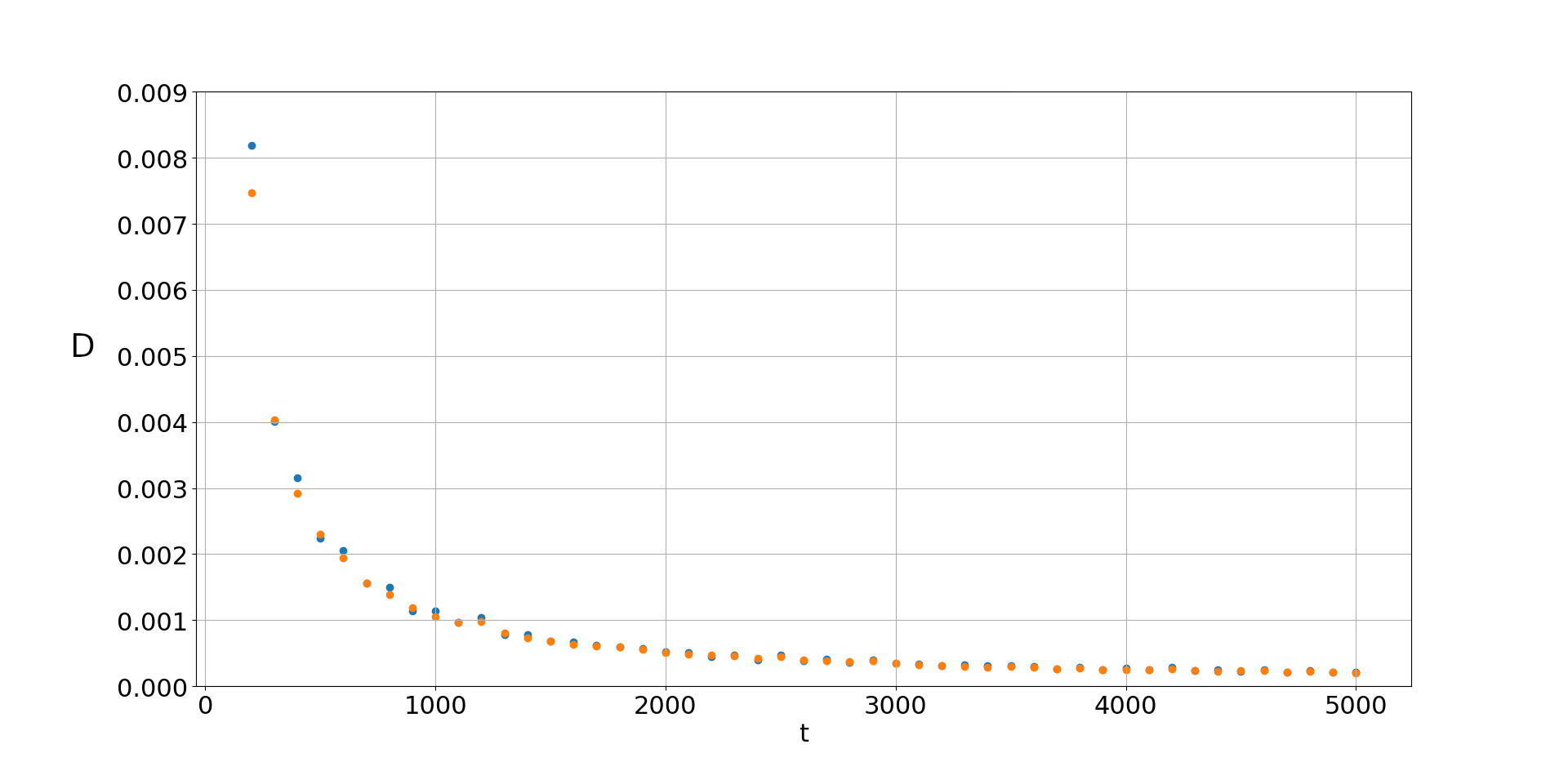}
\caption{Successive Frobenius norms for $c_2=0.2$ (blue) and $c_2=0.5$ (orange) when the initial distribution satisfies Born's rule. The orange dots cover most of blue dots.}\label{self_distance_ikanopoiisi}
\end{figure}

In Fig.~\ref{self_distance_ikanopoiisi} we calculate $D$ between the patterns at $t=0, 100, 200,\dots 5000$ for two initial distributions of 2400 particles which satisfy BR.  We see that $D$ is always smaller than $0.01$ and tends to zero as $t$ increases. In fact beyond $t=2000$ it is smaller than $D=0.0003$. In all the distributions of particles considered below we find that a final pattern is reached after a time $t=5000$, while in the case of individual trajectories a final pattern is reached after much larger times (of order $10^6$). 

\begin{figure}[H]
\centering
\includegraphics[scale=0.25]{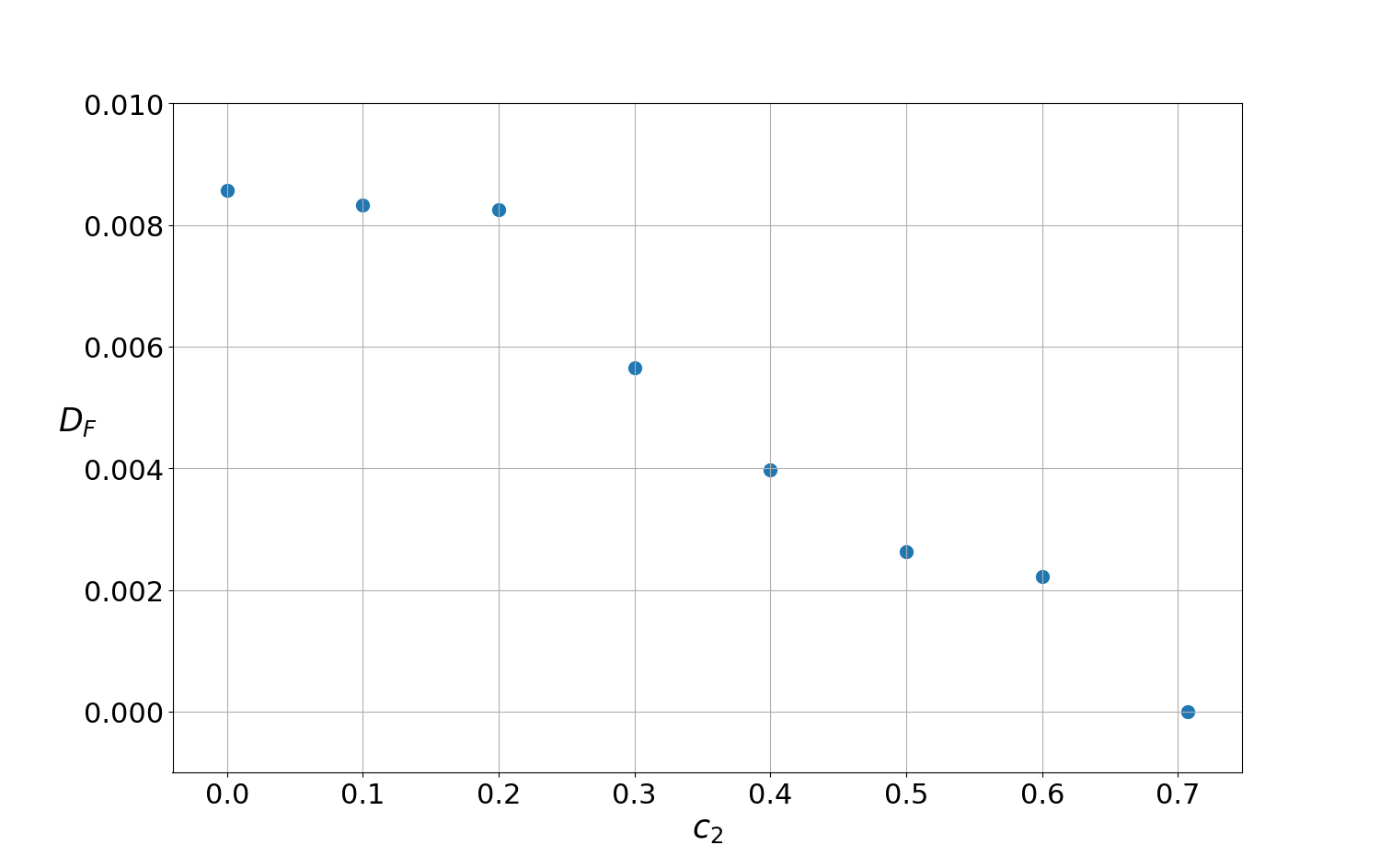}
\caption{The Frobenius norm  $D_F$ between  an initial distribution that satisfies Born's rule and the distribution of $c_2=\sqrt{2}/2$ that satisfies initially Born's rule, as function of the entanglement.}\label{0707alles}
\end{figure} 

Even though the two blobs $|\Psi|^2$ vary with $c_2$, following the changes of $|\Psi|^2$ discussed in Section 2, the final  patterns of the points of the trajectories for various $c_2$ are very similar.  In Fig.~\ref{0707alles} we compare the final patterns for various values of $c_2$ with that of the maximum entanglement $c_2=\sqrt{2}/2$ and  find a final Frobenius norm  $D_F=0.00262$ for $c_2=0.5$, $D_F=0.00825$ for $c_2=0.2$ and $D_F=0.00857$ for $c_2=0$. The values of $D_F$ increase as $c_2$ decreases, and their small values (smaller than $0.01$) account for the similarity  between the  color plots for various $c_2$ and that of  Fig.~\ref{0707_md_satisfaction}.

%
%
%

\begin{figure}[H]
\centering
\includegraphics[scale=0.25]{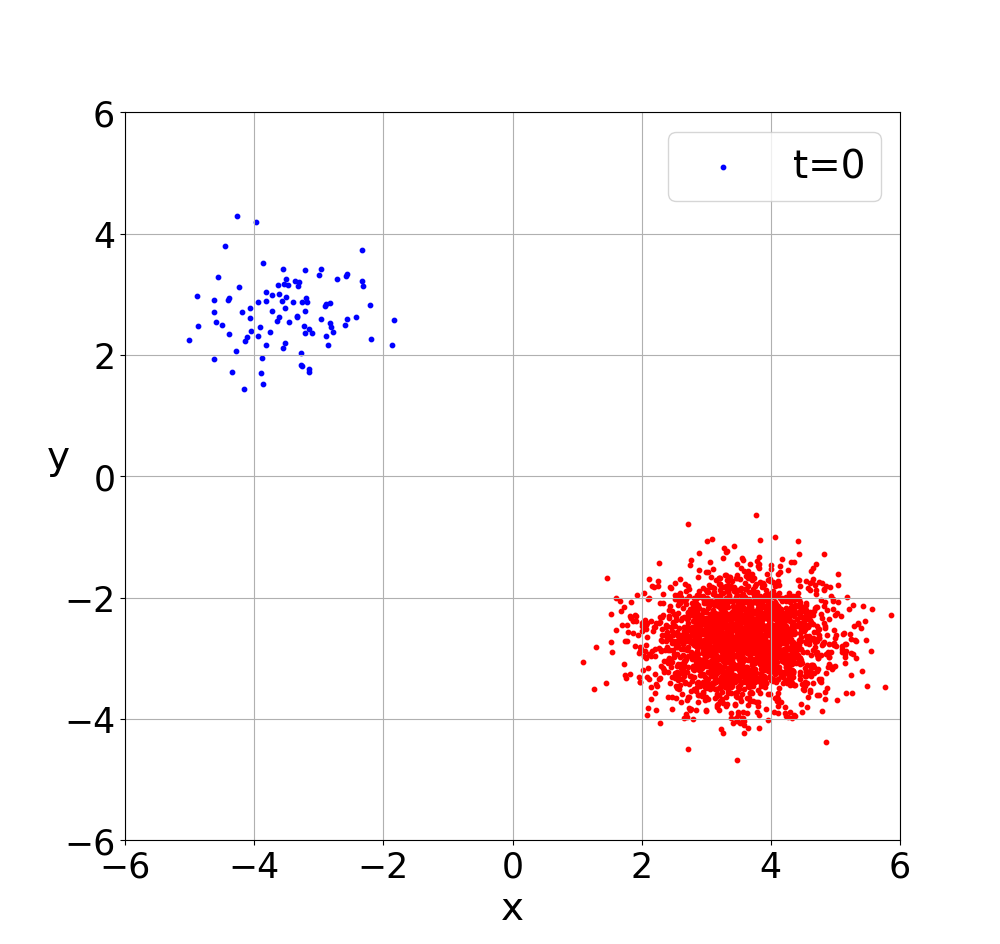}
\includegraphics[scale=0.25]{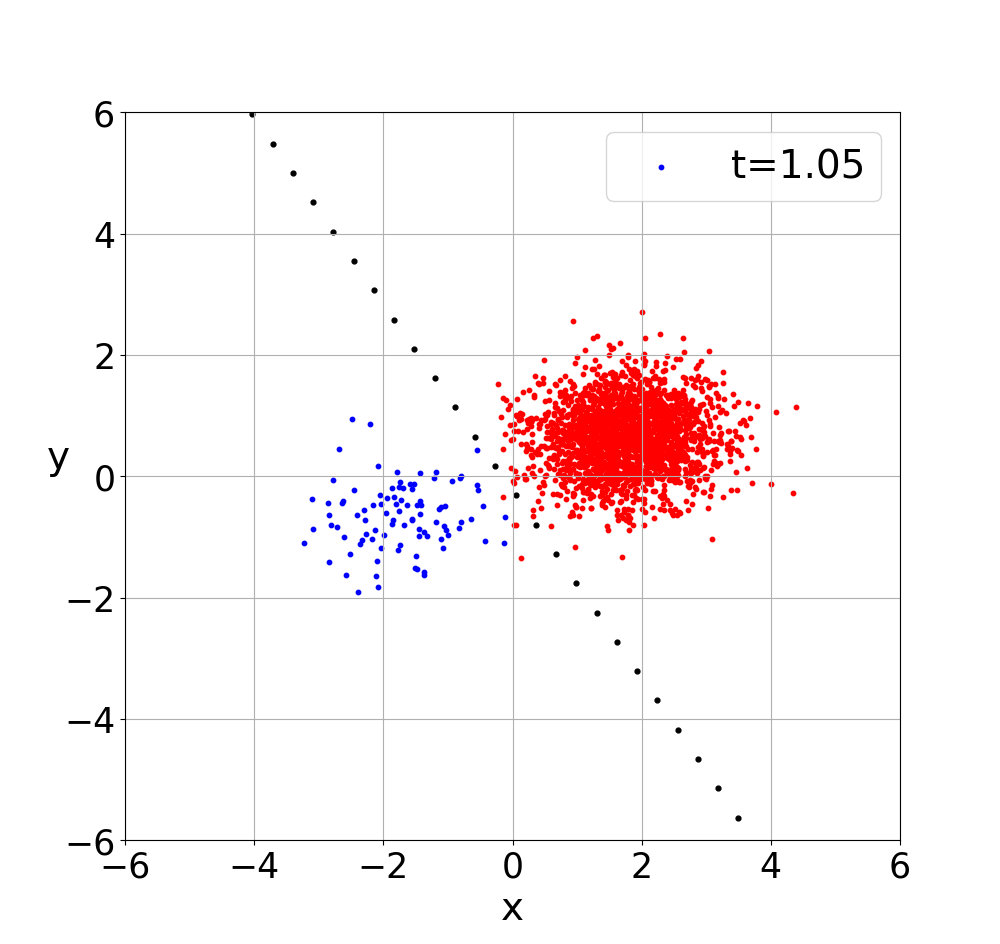}
\includegraphics[scale=0.25]{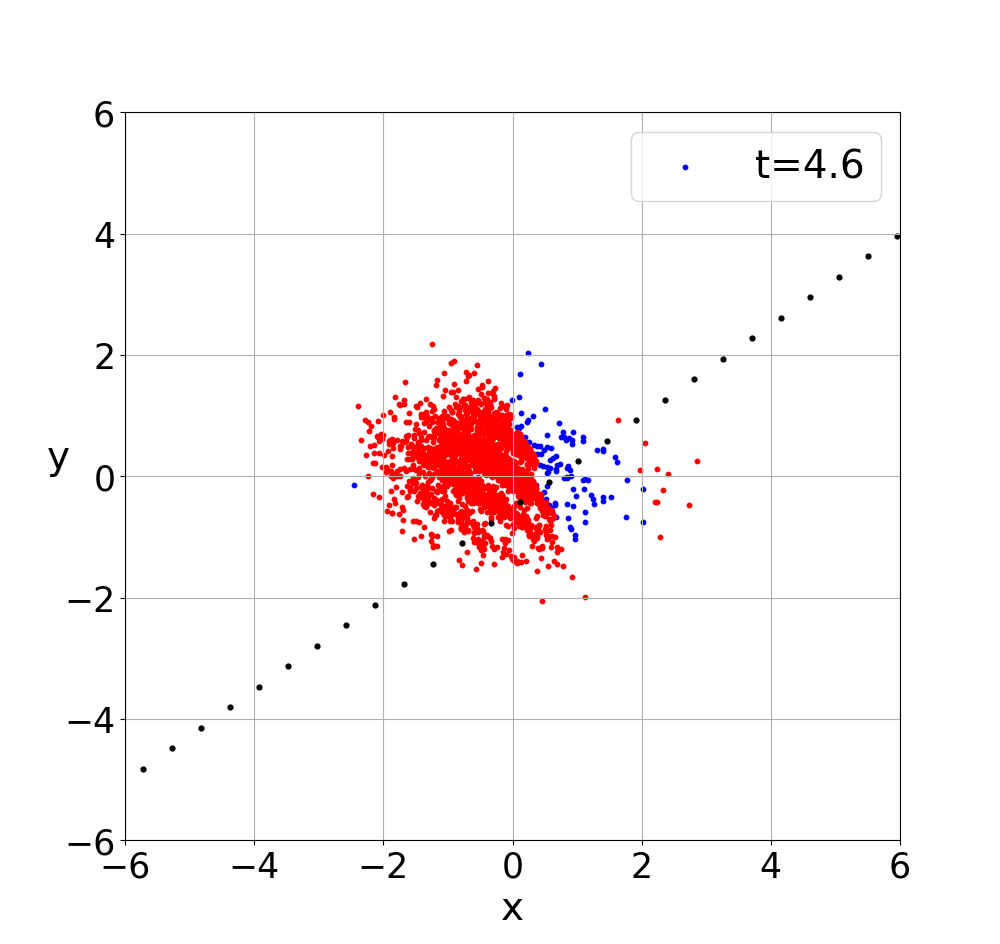}
\includegraphics[scale=0.25]{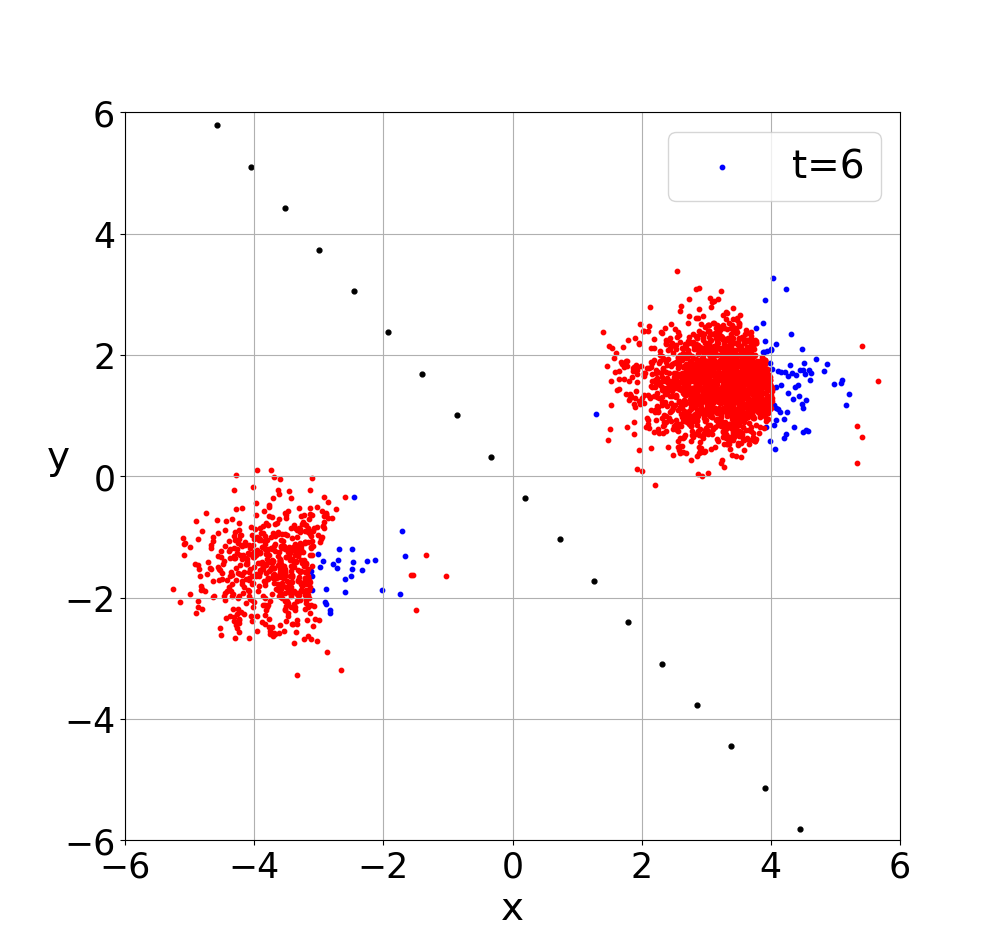}
\caption{Exhange of particles as two blobs aproach each other in the  case $c_2=0.2$  when Born's rule is initially satisfied with a total number of particles equal to 2400. (a) Initial conditions $t=0$ (b) approach $t=1.05$, (c) collision  $t=4.6$ (d) After the collision $t=6$ the blobs are formed again.}\label{distsbl}
\end{figure}

%
During the collisions several trajectories are deflected by approaching
one of the NPXPCs and they may go from one blob to the other. Nevertheless
the blobs are formed again after every collision  and they continue to satisfy
Born's rule $P=|\Psi|^2$. This was shown in Fig.~7 of our paper \cite{tzemos2020chaos} in the 
case of maximum entanglement. The same happens for other  values of the entanglement. E.g. in Figs.~\ref{distsbl}abcd we give the distributions of the points of the trajectories in the case of small
entanglement $c_2=0.2$ initially (Fig.~\ref{distsbl}a), at the first approach (Fig.~\ref{distsbl}b), at the first collision (Fig.~\ref{distsbl} c) and a little after this collision 
(Figs.~\ref{distsbl}d).  If the approach of the two blobs is not very close (Fig.~\ref{distsbl}b) only a few particles move from one blob to the other. If, however, we have a direct collision (Fig.~\ref{distsbl}c), many particles move to a different blob. However after the collisions the same blobs are formed again, although they are followed  by particles of different colors (Fig.~\ref{distsbl}d).  Then the points of the total set of trajectories form essentially the same overall picture as shown in Fig.~\ref{0707_md_satisfaction}.

\section{CHAOTIC VS ORDERED TRAJECTORIES}
In the case of zero entanglement, we have only one Gaussian blob (in the lower right part of the configuration space) and all the trajectories form  Lissajous figures (they are ordered) and each of them gives a different final pattern of points. Thus in this case the BR is satisfied only by an appropriate distribution of such figures. With a slight increase of the entanglement from zero, nodal points appear and generate chaotic trajectories in a large part of the  configuration space. The ordered trajectories are then confined near the center of the main blob of $|\Psi|^2$. In fact if $0<c_2<\sqrt{2}/2$ the main blob contains both ordered and chaotic trajectories, while the secondary blob contains only chaotic trajectories. In the limit of maximum entanglement ($c_2=\sqrt{2}/2$) the region of ordered trajectories disappears and the Born rule is always established, because all the trajectories are chaotic and ergodic. 
Consequently it is of great interest to understand when BR is accessible in the case of the partially entangled states.

\begin{figure}[H]
\centering
\includegraphics[scale=0.23]{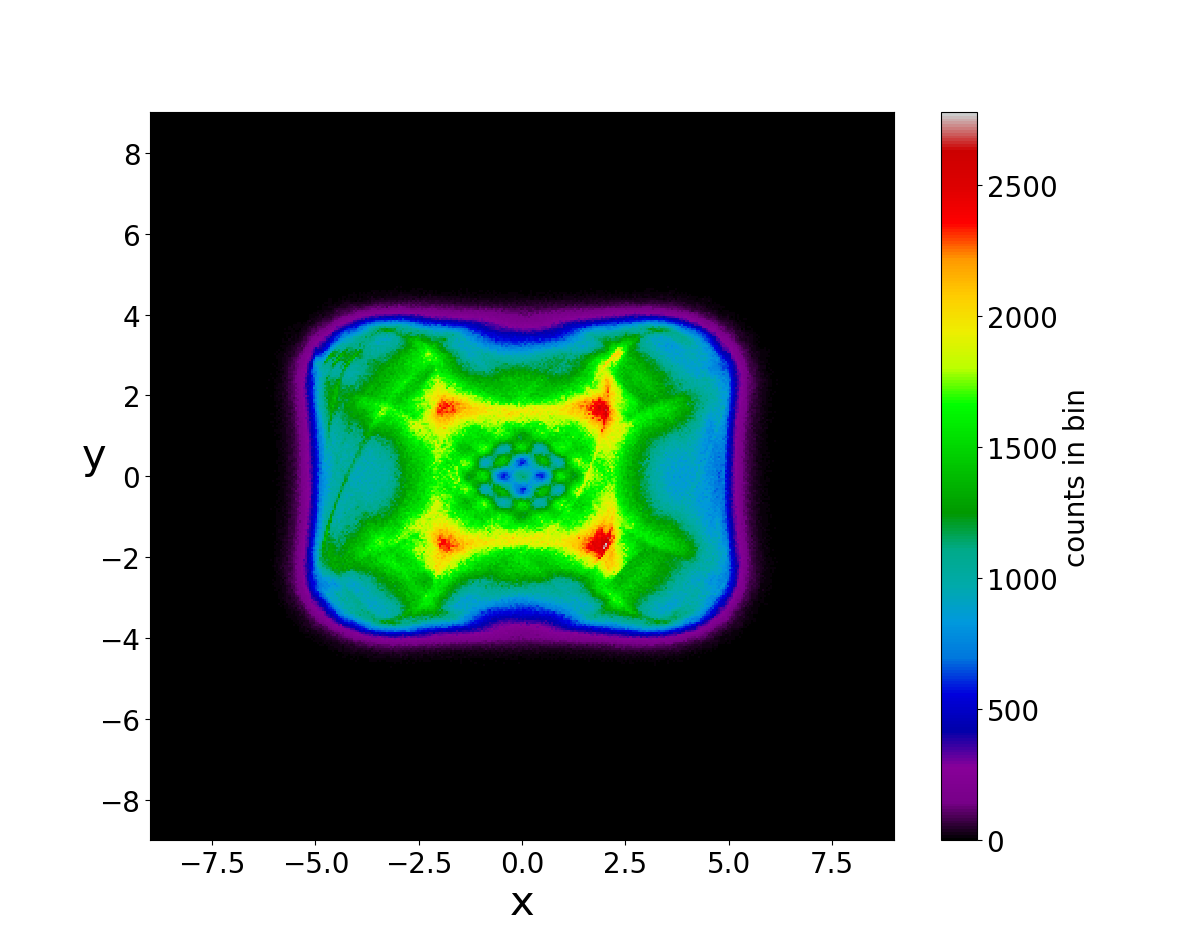}
\includegraphics[scale=0.23]{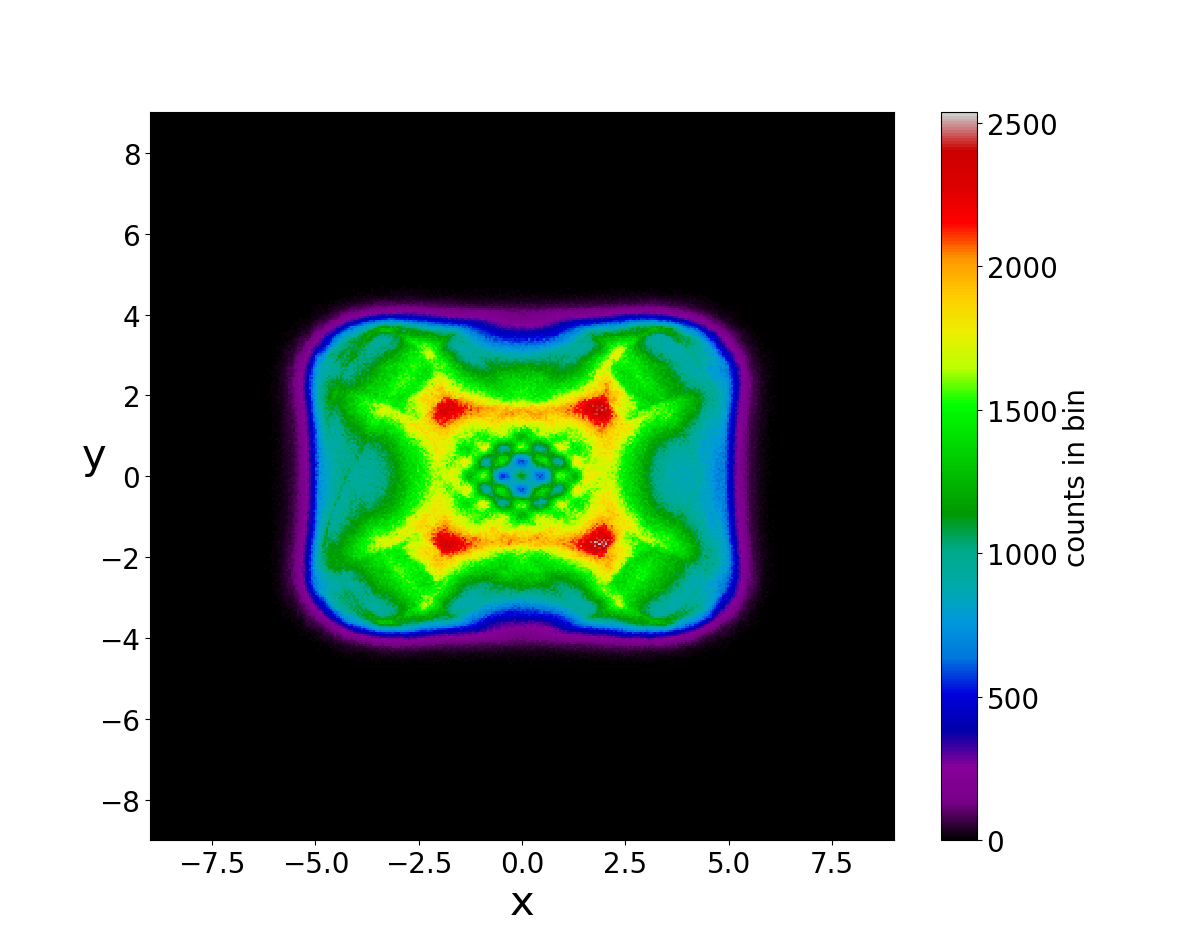}
\caption{Two  single chaotic-ergodic trajectories in the case $c_2=0.2$ for $t$ up to $2\times 10^6$. (a) $  x_0=-2.52027, y_0=2.17529$ and (b) $x_0=2, y_0=-2$.}\label{ekatommyria}
\end{figure}

We note again that he patterns of the points of individual chaotic  trajectories for every $c_2\neq 0$  are the same and it does not matter if a chaotic trajectory starts inside the main blob , or not. E.g. in Fig.~\ref{ekatommyria} we see the patterns of the points of two chaotic trajectories, one in the upper left and another one in the lower right (inside the main blob of $|\Psi|^2$). These patterns require a long time to be established, but the patterns found after a time $t=2\times 10^6$ are quite similar. 
We have found that the Frobenius norms between different chaotic trajectories for the same $c_2$ are smaller than $10^{-16}$. Therefore these trajectories are exactly ergodic.

\begin{figure}[H]
\centering
\includegraphics[scale=0.3]{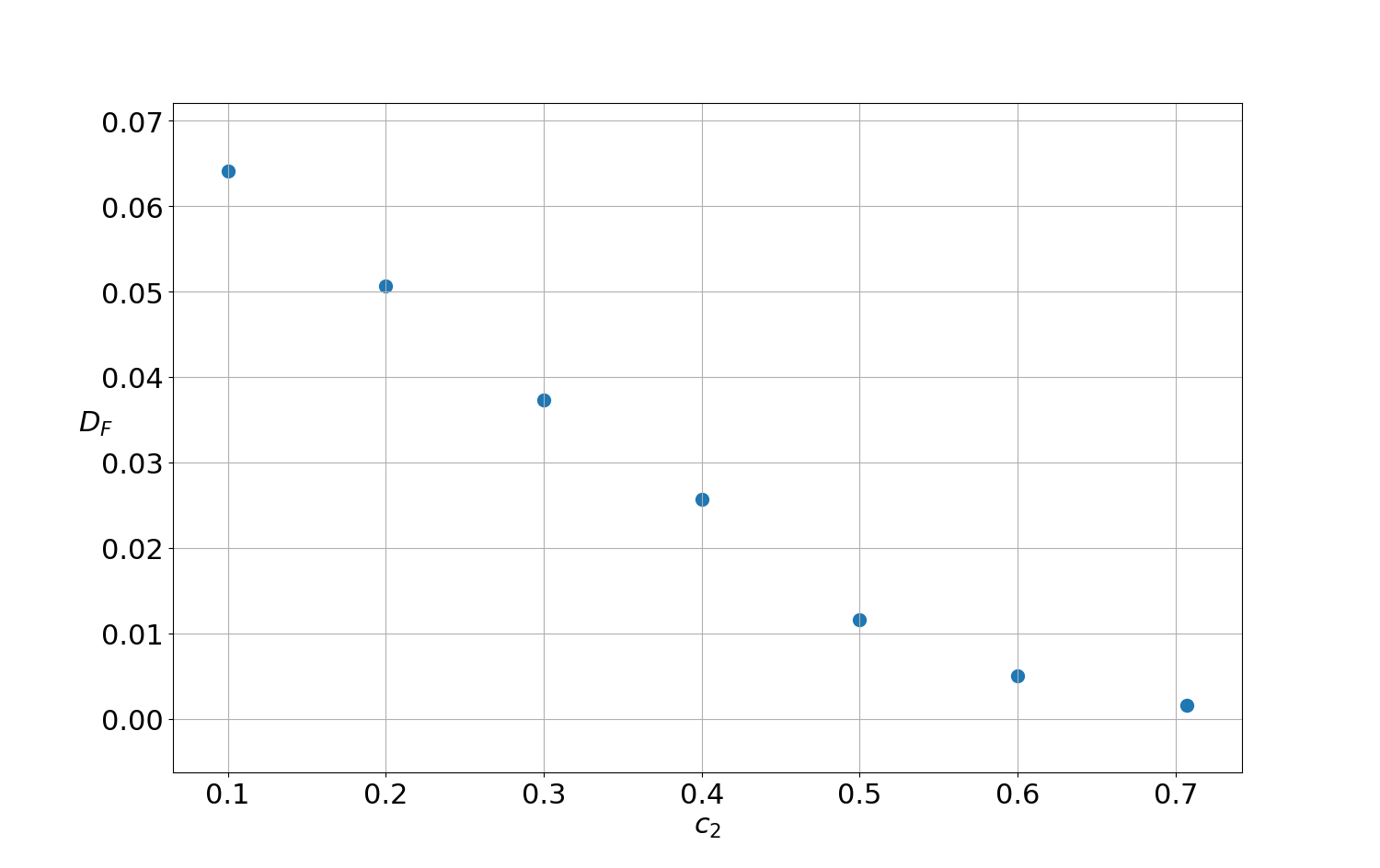}
\caption{The final Frobenius norm $D_F$ of the deviations of the patterns of the points of   individual chaotic trajectories from the corresponding Born patterns as a function of $c_2$. }\label{3chaotic}
\end{figure}

However the patterns of the points of the chaotic trajectories of different $c_2$ are different from the patterns that follow  Born's rule, due to the existence of ordered trajectories. Their difference increases as the value of $c_2$ decreases, as seen in Fig.~\ref{3chaotic}. We see that the  final Frobenius norm is about $D_F=0.01$ or smaller if $0.5<c_2<0.707$ and tends to zero as $c_2$ tends to $\sqrt{2}/{2}$.


In particular in the case $c_2=0.2$ we see that the pattern of Fig.~\ref{ekatommyria} forms 4 red spots at $x=\pm 2.4, y=\pm 1.8$, while the red spots in the case of Born's rule (close to the case of Fig.\ref{0707_md_satisfaction}) are at $x=3.0, y=\pm 2.2$. As $c_2$ becomes smaller than $c_2=0.1$ the deviations become even larger (see, e.g. the case $c_2=0.001$ in \cite{tzemos2020ergodicity}).

These differences  stem from the fact that for $c_2<\sqrt{2}/2$ there is a number of ordered trajectories in the lower right blob of Born's rule and this proportion increases as $c_2$ decreases.  The ordered trajectories are  deformed Lissajous curves and it is only their  collective pattern, together with the  collective pattern of the appropriate proportion of chaotic trajectories, that generates the Born rule after a long time.

\begin{figure}[H]
\centering
\includegraphics[scale=0.3]{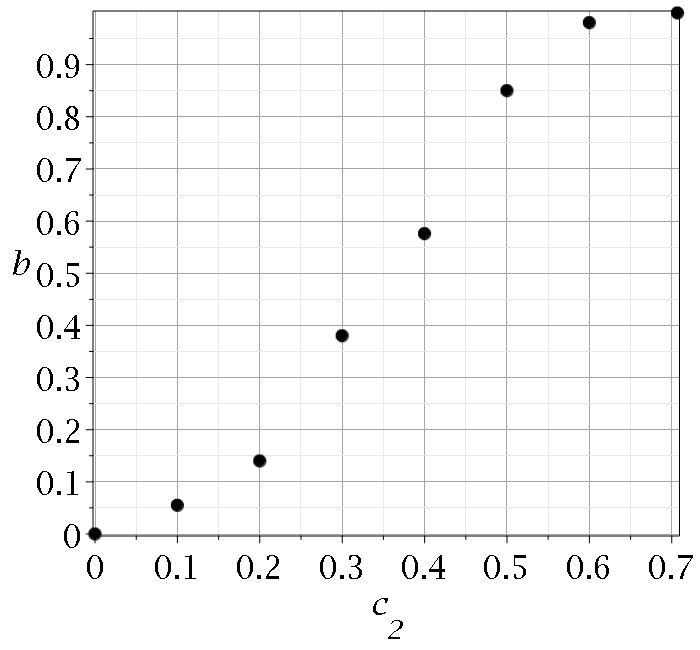}
\caption{The percentage $b$ of the chaotic trajectories on the main blob (lower right) of $|\Psi_0|^2$ as a function of the entanglement in the case of Born's distribution, according to our approximative algorithm described in the Appendix. We observe the two extreme cases $c_2=0$ where all trajectories are ordered and $c_2=\sqrt{2}/2$ where all trajectories are chaotic. }\label{produp}
\end{figure}

The proportion of the chaotic trajectories,  $b$, in the lower right blob of the initial Born distribution  for various values of $c_2$ is given in  Fig.~\ref{produp}. The distinction between ordered and chaotic trajectories was made by an approximate algorithm that is described in the Appendix. The proportion of ordered trajectories is equal to zero for $c_2=\sqrt{2}/2$ and it is equal to $1$ (i.e. $100\%$) if $c_2=0$.

If now we take a set of particles consisting of a proportion $p_1$ on the upper left blob and  $p_2=1-p_1$ on the lower right blob, the total proportion of chaotic trajectories is 
\begin{align}\label{pch}
P_{ch}=p_1+bp_2
\end{align}
while the proportion of ordered trajectories is 
\begin{align}\label{por}
P_{or}=(1-b)p_2
\end{align}
(with $P_{ch}+P_{or}=1$).
Thus we find that the ratio between chaotic and ordered trajectories is:
\begin{align}\label{logos}
\frac{P_{ch}}{P_{or}}=\frac{p_1/p_2+b}{1-b}
\end{align}
For every value of $c_2$ the proportions $p_1/p_2$ and $b$ are fixed, thus the ratio $P_{ch}/P_{or}$ is also fixed. E.g. for $c_2=0.2$ we have $p_1/p_2=0.04$ and $b=0.14$ therefore $P_{ch}/P_{or}\simeq 0.21$. Similarly in the case $c_2=0.5$ we find $P_{ch}/P_{or}\simeq 7.9$.

From Fig.~\ref{produp} we conclude that the proportion of ordered trajectories is small for relatively large entanglement (i.e. $c_2>0.5$). In these cases the final Frobenius norm $D_F$ is small (Fig.~\ref{3chaotic}). E.g. in the case $c_2=0.5$ the proportion of ordered trajectories is about $1-b=0.15$ and the $D_F$ is about $0.01$. Then the corresponding pattern of the points of the trajectories is quite close to that of BR.

\begin{figure}[H]
\centering
\includegraphics[scale=0.255]{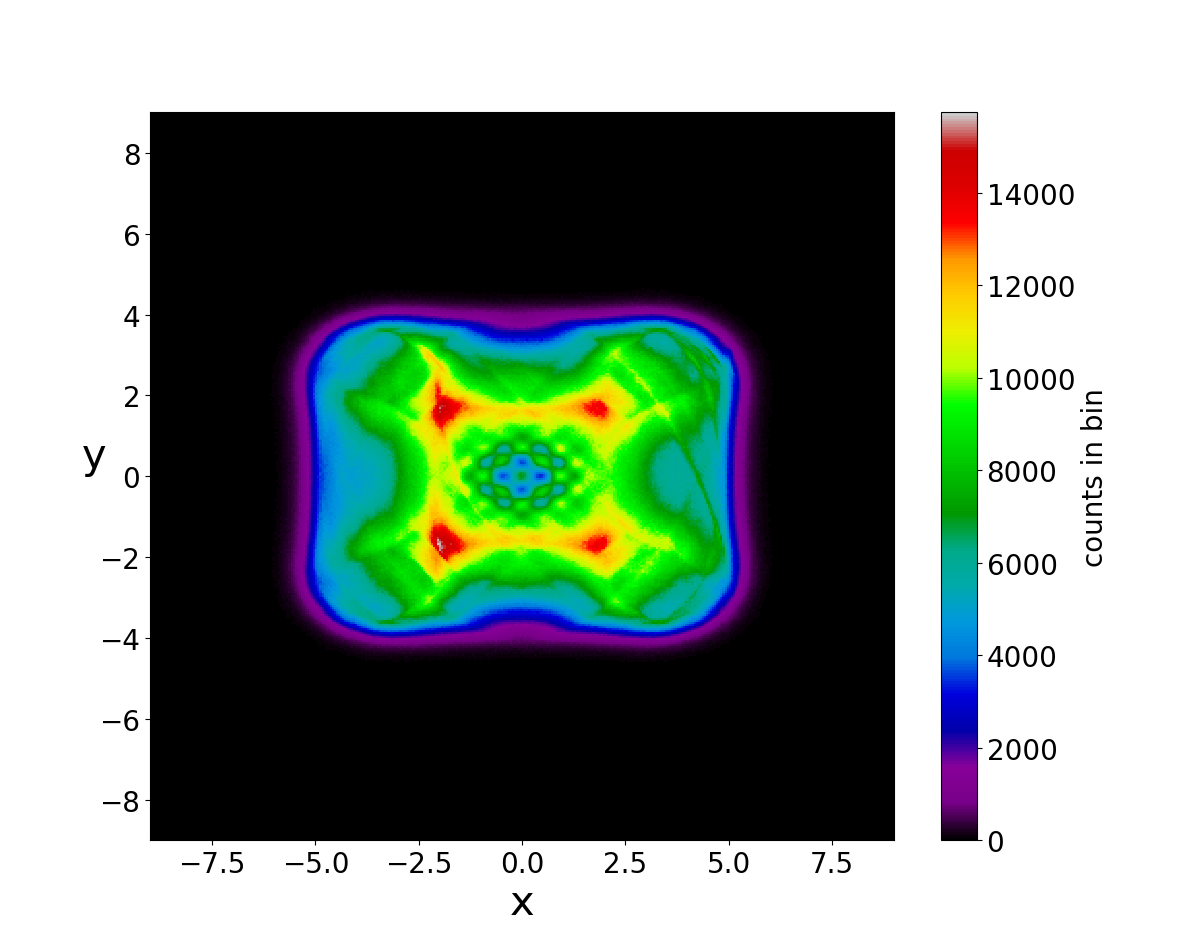}
\caption{The distribution of points of a particle distribution that violates initially Born's rule (2304 particles on the upper left blob and 96 on the lower right blob i.e. inverse proportions from the proportions of Born's rule) for $c_2=0.2$ and $t=5000$.}\label{02paraviasi}
\end{figure}

\begin{figure}[H]
\centering
\includegraphics[scale=0.25]{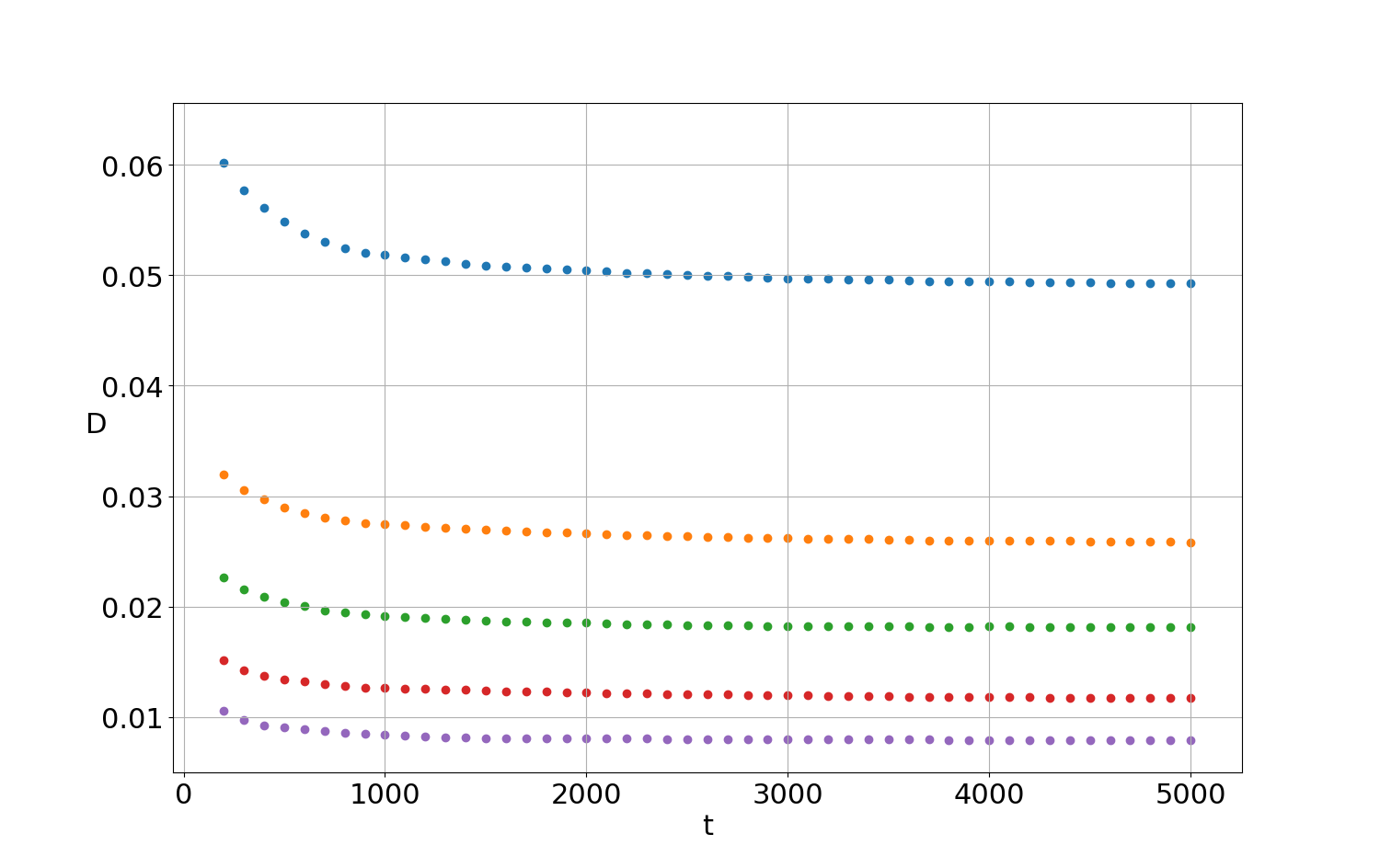}
\caption{ The Frobenius norm $D$ of the patterns of the points of the trajectories for $c_2=0.2$ between the Born distribution of particles at $t=5000$ and of  three initial distributions which violate Born's rule. The blue curve corresponds to 2304 particles on the upper left blob ($p_1=0.96$) and 96 on the lower right blob (proportion $p_2=0.04$), while the orange  and green curves correspond to 1200-1200 particles ($p_1=p_2=0.5$) and 800-1600 ($p_1=1/3, p_2=2/3$) in the two blobs. Finally the red and purple curves correspond to 500-1900 particles ($p_1=0.21, p_2=0.79$) and 200-2200 particles $(p_1=0.08, p_2=0.92)$ correspondingly.}\label{02_distance_from_borns_rule}
\end{figure}

On the other hand for weak entanglement (i.e. $c_2<0.3$) the proportion of ordered trajectories is relatively large. E.g. for $c_2=0.2$ the proportion of ordered trajectories is about $1-b=0.86$. Then  $D_F\simeq0.05$ if the initial distribution is 100\% around the upper left blob, and the pattern of the points of the trajectories differs significantly from that of BR, as seen in Fig.~\ref{02paraviasi}. In fact in Fig.~\ref{02paraviasi} is practically identical with the final pattern of the points of individual trajectories of Fig.~ \ref{ekatommyria}. Of course if we take a larger proportion of the initial conditions around the lower right blob the difference from Born's rule becomes smaller, as seen in Fig.~\ref{02_distance_from_borns_rule} and becomes zero when we take about 96\% in the lower left blob (the proportion of Born's rule itself).

If $c_2$ is even smaller (smaller than $c_2=0.1$) the deviations from BR are larger, and when $c_2=0$ they become maximum, unless of course we populate the lower right blob with the great majority (the totality if $c_2=0$) of initial conditions, as required by BR.

\begin{figure}[H]
\centering
\includegraphics[scale=0.15]{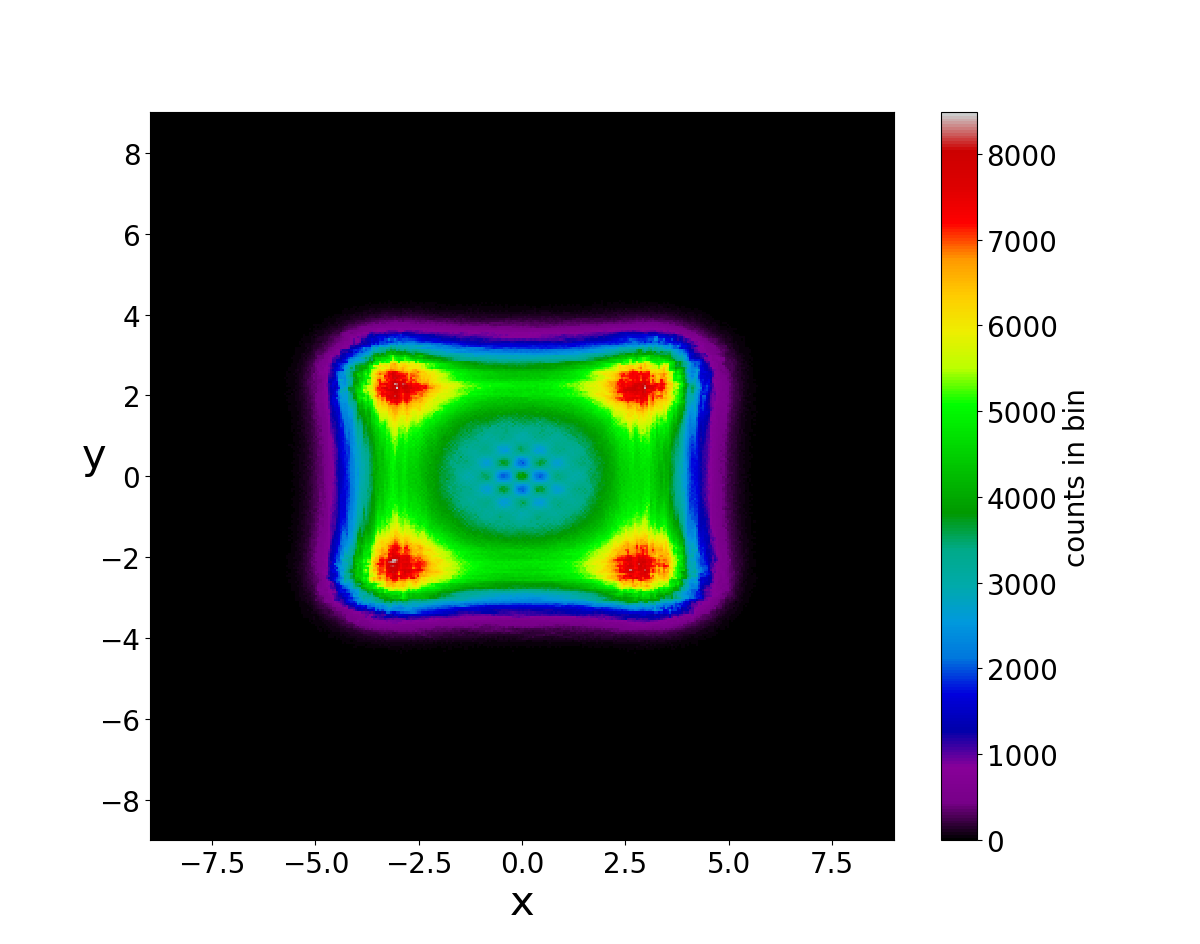}
\includegraphics[scale=0.21]{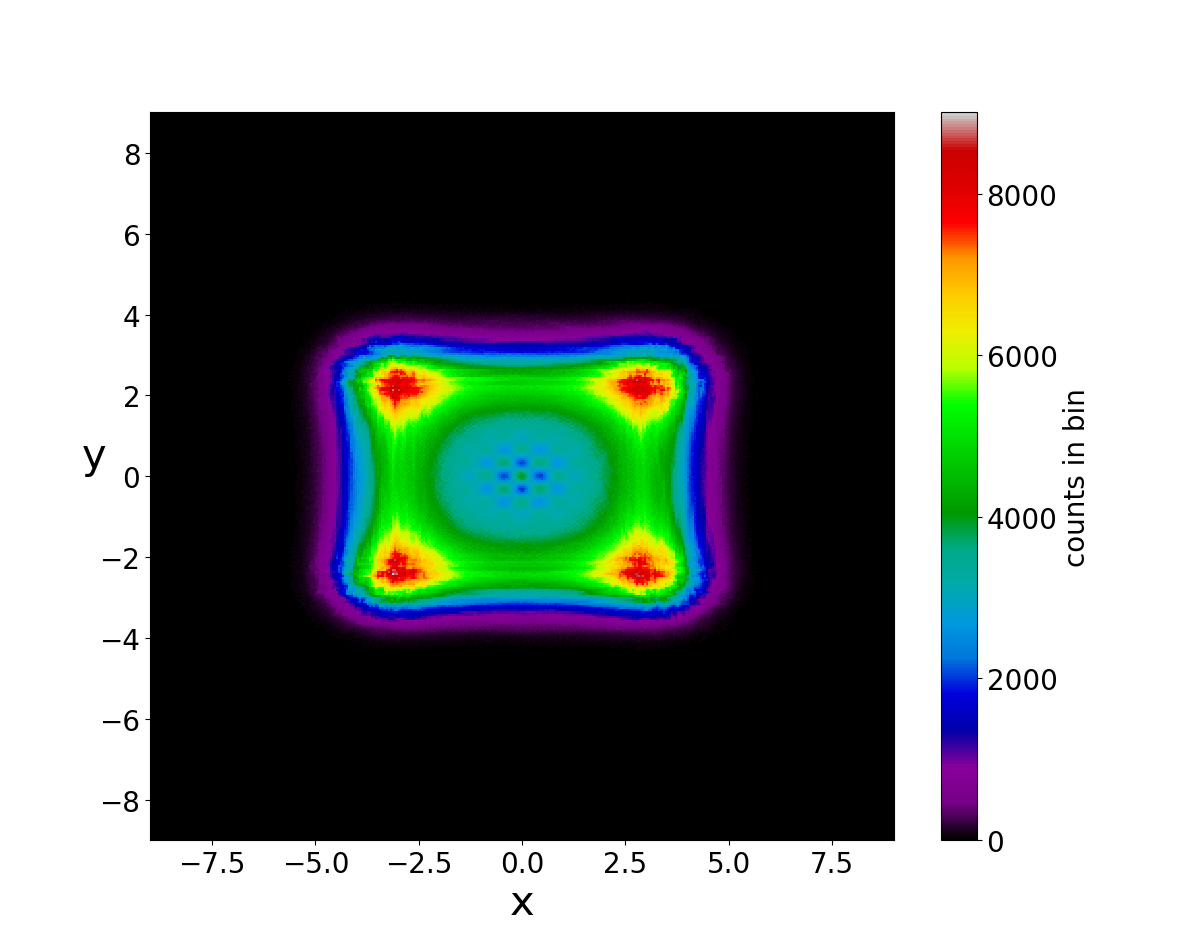}
\includegraphics[scale=0.21]{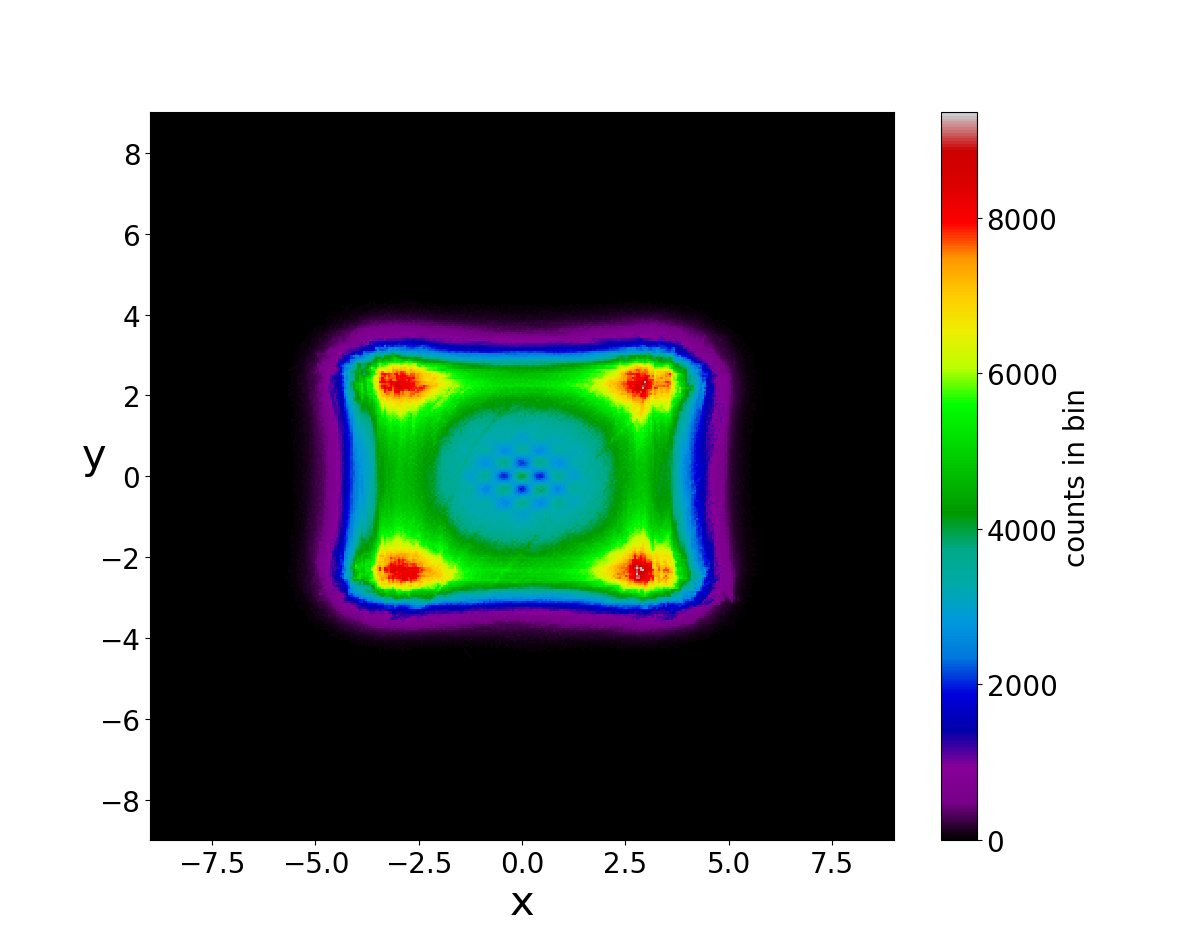}
\caption{A realization of Born's rule in the case $c_2=0.2$ with 1000 particles on the main blob and 40 on the upper left blob for $t=5000$. b) An initial violation of Born's rule with 1000 particles on the main blob and 40 on a blob  around the point $(-3.54, -1.69)$. c) An initial violation with 850 ordered trajectories on the main blob and 190 chaotic trajectories on a blob around the point $(3.54,-1)$. We observe the similarity of the three figures.}\label{analogy}
\end{figure}

However if we take the initial set of ordered trajectories, as we found in applying BR for a given $c_2$, we can take the remaining  set of chaotic trajectories anywhere and then we always recover the Born distribution in the long term. We have checked that in a number of cases by taking all the chaotic trajectories around the upper left blob or elsewhere. Three examples are given in Fig.~\ref{analogy}, where we compare an initial distribution satisfying BR (Fig.~\ref{analogy}a) with  distributions violating initially BR but with the same ratio between chaotic and ordered trajectories (Figs~\ref{analogy}bc). In particular in Fig.~\ref{analogy}c we have taken an initial violation of BR,  where the main blob has only ordered trajectories and all the chaotic trajectories are taken in another blob around the point $(3.54, -1.69)$, but with the same ratio $P_{ch}/P_{or}$. We observe their close similarity. Consequently it is the ratio between the ordered and chaotic-ergodic trajectories which makes BR accessible (or not).


\section{DISTRIBUTIONS OF TRAJECTORIES WITH $P_0\neq|\Psi_0|^2$}

If we take initial conditions of particles different from those of BR we may (or not) approach BR after a long time. As we have seen in the previous section Born's rule is reached for any initial distribution of particles in the case of maximum entanglement ($c_2=\sqrt{2}/2$). For smaller values of $c_2$ Born's rule is reached for any distribution of chaotic trajectories, provided that the proportion of ordered trajectories is the same with that of  Born's rule. If, however, the proportion of ordered trajectories is smaller (or larger) than that provided by BR rule we never recover Born's rule in the long run. E.g. this happens if we take  particles with initial conditions in the upper left and in the lower right blob with a ratio $p_1/p_2$  different from that of BR.  We have also a deviation of the ratio $P_{ch}/P_{ord}$ (according to Eq.~(\ref{logos})), therefore we cannot reach BR after a long time.

However, if the proportion of ordered trajectories is close to that required by Born's rule, then the deviation of the pattern of the trajectories from that of Born's rule is small.

\begin{figure}[H]
\centering
\includegraphics[scale=0.25]{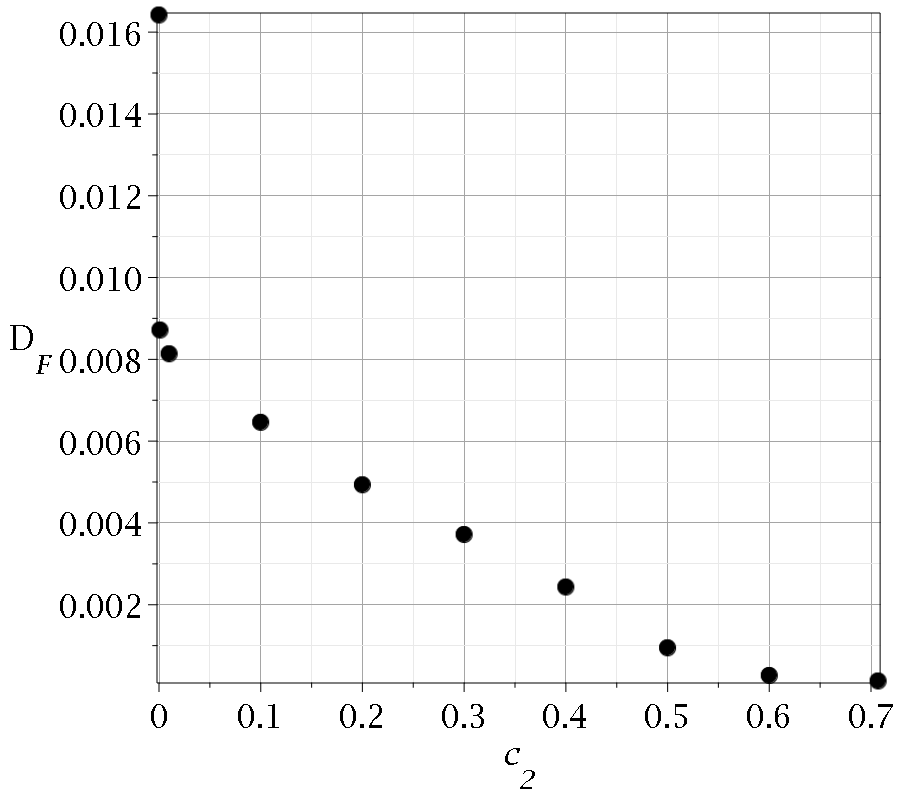}\\
\caption{Successive Frobenius norms comparing the  evolution of initial distributions lying $100\%$ in the upper left blob and the Born distribution at $t=5000$ for $c_2=0,0.001, 0.01, 0.1, 0.2 ,0.3 ,0.4 ,0.5, 0.6, 0.707$.}\label{allup}
\end{figure}

In order to find quantitatively the deviations from Born's rule we have considered two examples.  In the  first example we give the final Frobenius norm $D_F$ (deviations from Born's rule) for various values of $c_2$ when the initial distribution of particles is 100\% in the upper left blob (Fig.\ref{allup}). We see that when $c_2=0$ this norm is relatively large ($D_F\simeq 0.165$), but when $c_2$ increases this norm decreases considerably and for $c_2\geq 0.5$ it is smaller than $D_F=0.01$. This means that for relatively large $c_2$ the final pattern is very close to that of  BR. 

\begin{figure}[H]
\centering
\includegraphics[scale=0.4]{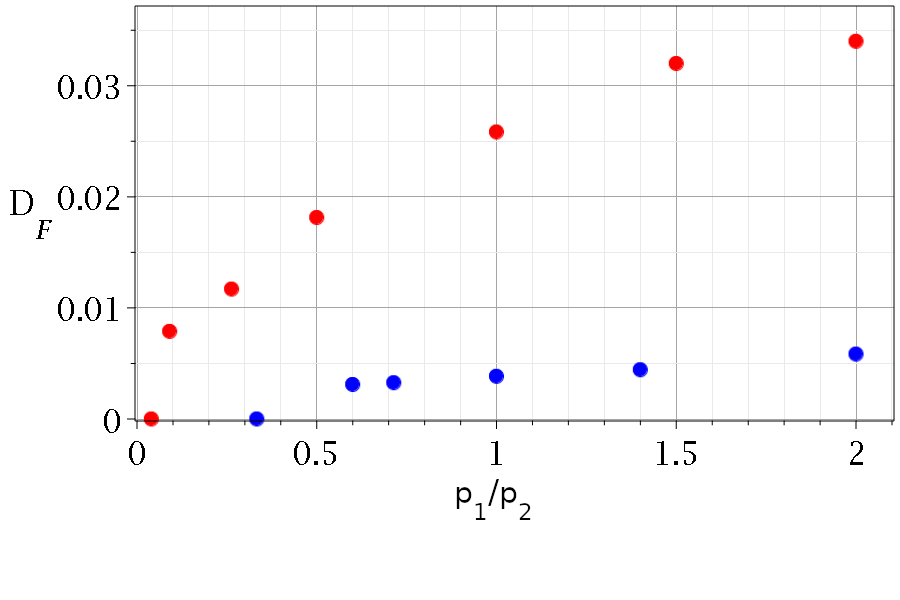}
\caption{The final Frobenius norm $D_F$ as a function of the proportion $p_1/p_2$ of the initial particles in the upper left blob and in the lower right blob for  $c_2=0.2$ (red dots) and $c_2=0.5$ (blue dots).}\label{Nomos}
\end{figure}

In the second example we calculated the final Frobenius norm $D_F$ of the deviations of the final pattern of the points of the  trajectories for various proportions $p_1/p_2$ of initial conditions in the upper left and in the lower right blob in the cases $c_2=0.2$ and $c_2=0.5$ (Fig.~\ref{Nomos}).  We see that as $p_1/p_2$ decreases the values of $D_F$ become smaller. When the ratio $p_1/p_2$ tends to the value appropriate for Born's rule the value of $D_F$ tends to zero. However for smaller $p_1/p_2$ the values of $D_F$ become again positive. In the case $c_2=0.5$ we have $D_F\leq 0.005$ for $p_1/p_2<2$, therefore we find again that for large $c_2$ ($c_2\geq 0.5$) the final pattern is  very close to BR. On the other hand for $c_2=0.2$ we have $D_F>0.025$ for $p_1/p_2>1$, therefore the final deviation from BR is larger and only if $p_1/p_2$ is smaller than $0.2$ we have $D_F<0.01$, i.e. we come close to Born's rule.   Therefore for $c_2=0.2$ or less BR is not satisfied in general.



%
%

\section{Conclusions}
In the present paper we studied the role of chaotic and ordered trajectories in establishing Born's rule,  in a paradigmatic entangled 2-qubit system.


We  calculated many trajectories for various values of $c_2$ and  found the patterns of their points over the course of time.  We established the following:
\begin{enumerate}
\item The form of $|\Psi|^2$ generates two blobs for various values of the entanglement, one on the lower right from the origin (main blob) and the other on the upper left from the origin. The two blobs approach each other from time to time and undergo several collisions, where we have the formation of secondary blobs. The collisions occur at practically the same times for all the values of the entanglement. After the collisions the two blobs are formed again.
\item If the initial distribution $P_0$ satisfies Born's rule $P_0=|\Psi_0|^2$ then it is known that this distribution follows the evolutuion of $|\Psi|^2$ for all times. During the collisions the two blobs exchange particles and later on the blobs consist of a mixture of particles from the initial blobs.
\item The exchanges of particles occur when particles approach the nodal points, where $\Psi=0$, and the nearby X-points. There is an infinite number of nodal points along a straight line, where the distances between the nearby nodal points are the same for any value of $c_2$, but they change in time. These distances are minimal during the collisions.


\item The differences between successive in time  patterns, giving the distribution of the points of the trajectories, decrease, as time increases,  and tend to zero, giving a final norm $D_F$ for every value of the entanglement.

\item The difference between the final Born pattern and the final pattern of the maximum entanglement case is small for any amount of the entanglement (the final Frobenius norm $D_F$ is less than $0.01$)  and decreases as the  entanglement increases.

\item The main blob of Born's rule consists of chaotic and ordered trajectories. Ordered trajectories appear near the center of the main blob. The proportion of ordered trajectories increases as the entanglement decreases. When the entanglement is maximum all the trajectories are chaotic and when the entanglement is zero  all the trajectories are ordered. The initial secondary blob (upper left) consists practically of only chaotic trajectories.

\item For any given value of entanglement the points of the individual chaotic trajectories form the same pattern. The differences between the patterns of various chaotic trajectories are insignificant (The Frobenius norm of their differences is smaller than $10^{-16}$).  Consequently the chaotic trajectories are always ergodic. 

\item If we take the proportion of ordered trajectories for a given amount of entanglement, according to Born's rule, then for any initial distribution of the chaotic trajectories the final pattern of the points of the trajectories tends to that of BR.
\item However, if the ratio between chaotic and ordered trajectories is different from that of  BR, then  the final pattern of the points of the trajectories is also different from that of BR and the difference increases as the entanglement decreases. The difference is small for strongly entangled states and it is large for  weakly entangled states. The difference is also small for any value of entanglement if the initial proportions $p_1$ and $p_2$ of particles in the upper left blob and in the lower right blob have a ratio $p_1/p_2$ close to the ratio of the Born rule.




\end{enumerate}

\section{Appendix}

Our numerical results show clearly the key role of the ratio between the chaotic and ordered trajectories for the approach of an arbitrary initial distribution to that of Born's rule. Consequenlty it is of fundamental importance to separate the ordered from the chaotic trajectories of an initial distribution. The standard way of doing this is to calculate the Lyapunov characteristic number (LCN) 
\begin{equation}
LCN=\lim_{t\to\infty}\chi,
\end{equation}
where $\chi$ is the `finite time LCN'
\begin{equation}\label{chi}
\chi=\ln(\xi/\xi_0)/t,
\end{equation} ($\xi_0, \xi$ are infinitesimal deviations $\xi(t)=\sqrt{\delta x^2+\delta y^2}$ at 
times $t_0=0$ and $t$). If LCN is a positive number then the trajectory is  chaotic and if LCN is zero then the trajectory is ordered. This method was followed in our previous works in these series of studies. However the calculation of LCN is a demanding computational problem, and would require a huge amount of work in this case where we focus on multiparticle distributions rather than single Bohmian trajectories.

In this work we avoided the calculation of thousands of LCNs
by exploiting the shape of the ordered trajectories of this model. As we have already seen the ordered trajectories are perfect or distorted Lissajous curves. Moreover in our previous work we showed that every Lissajous curve starts at its lower right corner and consequently its motion points initially to smaller $x$ and larger $y$. Finally the size of the perfect Lissajous curves (in the case $c_2=0$) is easily found and it is equal to
\begin{align}
|\Delta x_{max}|=\frac{2a_0\sqrt{2}}{\sqrt{\omega_x}},\quad |\Delta y_{max}|=\frac{2a_0\sqrt{2}}{\sqrt{\omega_y}}
\end{align}  
Consequently if we calculate the trajectories of a  distribution of $N$ particles for a quite long time, we can ask how many of them have exceeded significantly the area of the Lissajous curve plus a sufficient amount of space to  larger  and lower $x$ and $y$ than those at $t=0$ in order to cover the case of distorted Lissajous curves. These are characterized as chaotic curves.

This method is of course just an approximation but in the limit of large $t$ and $N$ it gives reliable results. The results of this method for the lower right blob of the $|\Psi|^2$  with $N=2400$ particles and $t=10^3$ are shown in Fig.~\ref{produp}.

\section*{References}
\bibliographystyle{iopart-num}
\bibliography{bibliography}

\end{document}